%% file: paper.tex
\documentclass[twoside,leqno,twocolumn]{article}

\usepackage[letterpaper]{geometry}

\usepackage{ltexpprt}
\usepackage{hyperref}
\usepackage{amsmath,amssymb,amsfonts}
\usepackage{graphicx}
\usepackage{textcomp}
\usepackage{xcolor}

\usepackage[ruled, vlined, nofillcomment, linesnumbered]{algorithm2e}
\usepackage{url}
\usepackage{amsmath}
\usepackage{mathtools}
\usepackage{graphicx}
\usepackage{subcaption}
\usepackage[english]{babel}
\usepackage{booktabs}
\usepackage{verbatim}
\usepackage{float}
\usepackage{footmisc}
\usepackage{xspace}
\usepackage{tikz}
\usepackage{pgfplots}
\usepackage{pgfplotstable}
\usetikzlibrary{shapes.geometric}
\usepackage[numbers,sort&compress]{natbib}

\usepackage{etoolbox}

\makeatletter
\patchcmd{\@algocf@start}
  {-1.5em}
  {0pt}
  {}{}
\makeatother
\input{defines}

\begin{document}

\title{Node ranking in labeled networks\thanks{This research is supported by the Academy of Finland projects MALSOME (343045)}}

\author{Chamalee Wickrama Arachchi, Nikolaj Tatti~\thanks{HIIT, Helsinki University, firstname.lastname@helsinki.fi}}
\date{}

\maketitle

\begin{abstract}\small\baselineskip=9pt
The entities in directed networks arising from real-world interactions
are often naturally organized under some hierarchical structure.
Given a directed, weighted, graph with edges and node labels, we introduce
ranking problem where the obtained hierarchy should be described
using node labels. Such method has the advantage to not only rank the nodes
but also provide an explanation for such ranking.
To this end, we define a binary tree called label tree, where each leaf represents a rank and each non-leaf
contains a single label, which is then used to partition, and consequently, rank the nodes in the input graph.
We measure the quality of trees using agony score, a penalty score that
penalizes the edges from higher ranks to lower ranks based on the severity of
the violation.
We show that the problem is \np-hard, and even inapproximable if we limit the size of the label tree.
Therefore, we resort to heuristics, and design a divide-and-conquer algorithm which runs in   $\bigO{(n + m) \log n + \ell R}$, where $R$ is the number of node-label pairs in the given graph, $\ell$ is the number of nodes in the resulting label tree, and $n$ and  $m$ denote the number of nodes and edges respectively.
We also report an experimental study that shows that our algorithm 
can be applied to large networks, that it can find ground truth in synthetic datasets,
and can produce explainable hierarchies in real-world datasets.
\end{abstract}

\input{intro}
\input{prel}
\input{algorithm}
\input{related}
\input{exp}
\input{conclusions}

\bibliographystyle{abbrvnat}
\bibliography{bibliography}

\input{appendix}
\end{document}

%% file: defines.tex
\SetKw{Output}{output}

\newcommand{\set}[1]{\left\{#1\right\}}

\newcommand{\fpr}[1]{\mathopen{}\left(#1\right)}

\newcommand{\abs}[1]{{\left|#1\right|}}

\newcommand{\np}{\textbf{NP}}
\newcommand{\apx}{\textbf{APX}}
\newcommand{\poly}{\textbf{P}}

\newcommand{\define}{\leftarrow}

\newcommand{\fm}[1]{{\mathcal{#1}}}

\DeclareRobustCommand{\dispfunc}[2]{%
	\ensuremath{%
		\ifthenelse{\equal{#2}{}}%
			{\mathit{#1}}%
			{\mathit{#1}\fpr{#2}}}}

\newcommand{\score}[1]{\dispfunc{q}{#1}}
\newcommand{\labelscore}[1]{\dispfunc{q}{#1}}
\newcommand{\gain}[1]{\dispfunc{\Delta}{#1}}
\newcommand{\bigO}[1]{\dispfunc{\mathcal{O}}{#1}}

\newcommand{\diff}[1]{\dispfunc{d}{#1}}

\newcommand{\dtname}[1]{\textsl{#1}}

\newcommand{\algsplit}{\textsc{Test}\xspace}
\newcommand{\algleft}{\textsc{Construct}\xspace}

\newcommand{\algpartition}{\textsc{Greedy}\xspace}
\newcommand{\algexact}{\textsc{Agony}\xspace}

\newcommand{\fasprb}{\textsc{FAS}\xspace}

\newcommand{\opt}[1]{\dispfunc{opt}{#1}}

\newcommand{\labeldic}[1]{\dispfunc{V}{#1}}

\newcommand{\inback}[1]{\dispfunc{ib}{#1}}
\newcommand{\outback}[1]{\dispfunc{ob}{#1}}

\newcommand{\back}[1]{\dispfunc{b}{#1}}

\newcommand{\pen}[1]{\dispfunc{p}{#1}}

\newtheorem{definition}{Definition}[section]
\newtheorem{problem}{Problem}[section]

\newcommand{\prblagy}{\textsc{L-agony}\xspace}
\newcommand{\prbcover}{\textsc{Cover}\xspace}

\SetKw{Break}{break}
\SetKw{AND}{and}
\SetKw{OR}{or}

\pgfdeclarelayer{background}
\pgfdeclarelayer{foreground}
\pgfsetlayers{background,main,foreground}

\definecolor{yafaxiscolor}{rgb}{0.3, 0.3, 0.3}

\definecolor{yafcolor1}{rgb}{0.4, 0.165, 0.553}
\definecolor{yafcolor2}{rgb}{0.949, 0.482, 0.216}
\definecolor{yafcolor3}{rgb}{0.47, 0.549, 0.306}
\definecolor{yafcolor4}{rgb}{0.925, 0.165, 0.224}
\definecolor{yafcolor5}{rgb}{0.141, 0.345, 0.643}
\definecolor{yafcolor6}{rgb}{0.965, 0.933, 0.267}
\definecolor{yafcolor7}{rgb}{0.627, 0.118, 0.165}
\definecolor{yafcolor8}{rgb}{0.878, 0.475, 0.686}

\newlength{\yafaxispad}
\newlength{\yaftlpad}
\newlength{\yaflabelpad}
\newlength{\yafaxiswidth}
\newlength{\yafticklen}
\makeatletter
\def\pgfplots@drawtickgridlines@INSTALLCLIP@onorientedsurf#1{}
\makeatother

\newcommand{\yafdrawaxis}[4]{
	\pgfplotstransformcoordinatex{#1}\let\xmincoord=\pgfmathresult 
	\pgfplotstransformcoordinatex{#2}\let\xmaxcoord=\pgfmathresult 
	\pgfplotstransformcoordinatey{#3}\let\ymincoord=\pgfmathresult 
	\pgfplotstransformcoordinatey{#4}\let\ymaxcoord=\pgfmathresult 
	\pgfsetlinewidth{\yafaxiswidth} 
	\pgfsetcolor{yafaxiscolor}
	\pgfpathmoveto{\pgfpointadd{\pgfpointadd{\pgfplotspointrelaxisxy{0}{0}}{\pgfqpointxy{\xmincoord}{0}}}{\pgfqpoint{-0.5\yafaxiswidth}{\yafaxispad}}}
	\pgfpathlineto{\pgfpointadd{\pgfpointadd{\pgfplotspointrelaxisxy{0}{0}}{\pgfqpointxy{\xmaxcoord}{0}}}{\pgfqpoint{0.5\yafaxiswidth}{\yafaxispad}}}
	\pgfpathmoveto{\pgfpointadd{\pgfpointadd{\pgfplotspointrelaxisxy{0}{0}}{\pgfqpointxy{0}{\ymincoord}}}{\pgfqpoint{\yafaxispad}{-0.5\yafaxiswidth}}}
	\pgfpathlineto{\pgfpointadd{\pgfpointadd{\pgfplotspointrelaxisxy{0}{0}}{\pgfqpointxy{0}{\ymaxcoord}}}{\pgfqpoint{\yafaxispad}{0.5\yafaxiswidth}}}
	\pgfusepath{stroke}
}

\pgfplotscreateplotcyclelist{yaf}{%
{yafcolor5,mark options={scale=0.75},mark=o}, 
{yafcolor2,mark options={scale=0.75},mark=square},
{yafcolor3,mark options={scale=0.75},mark=triangle},
{yafcolor4,mark options={scale=0.75},mark=o},
{yafcolor1,mark options={scale=0.75},mark=o},
{yafcolor6,mark options={scale=0.75},mark=o},
{yafcolor7,mark options={scale=0.75},mark=o},
{yafcolor8,mark options={scale=0.75},mark=o}} 

\pgfkeys{/pgf/number format/.cd,1000 sep={\,}}

\pgfplotsset{axis y line=left, axis x line=bottom,
	tick align=outside,
	tickwidth=\yafticklen,
	clip = false,
    x axis line style= {-, line width = 0pt, color=black!0},
    y axis line style= {-, line width = 0pt, color=black!0},
    x tick style= {line width = \yafaxiswidth, color=yafaxiscolor, yshift = \yafaxispad},
    y tick style= {line width = \yafaxiswidth, color=yafaxiscolor, xshift = \yafaxispad},
    x tick label style = {font=\small, yshift = \yaftlpad, inner xsep = 0pt},
    y tick label style = {font=\small, xshift = \yaftlpad},
    every axis y label/.style = {at = {(ticklabel cs:0.5)}, rotate=90, anchor=center, font=\small, yshift = -\yaflabelpad, inner sep = 0pt},
    every axis x label/.style = {at = {(ticklabel cs:0.5)}, anchor=center, font=\small, yshift = \yaflabelpad},
    x tick label style = {font=\small, yshift = 1pt},
    grid = major,
    major grid style  = {dash pattern = on 1pt off 3 pt},
	every axis plot post/.append style= {line width=\yafaxiswidth} ,
	legend cell align = left,
	legend style = {inner sep = 1pt, cells = {font=\scriptsize}},
	legend image code/.code={%
		\draw[mark repeat=2,mark phase=2,#1] 
		plot coordinates { (0cm,0cm) (0.15cm,0cm) (0.3cm,0cm) };%
	} 
}

\newcommand{\pgfprintduration}[1]{%
	\ifthenelse{\equal{#1}{}}{---}{%
	\pgfmathsetmacro{\minutes}{floor(#1 / 60)}%
	\pgfmathsetmacro{\seconds}{#1 - 60*\minutes}%
	\pgfmathifthenelse{\minutes > 0}{"\pgfmathprintnumber{\minutes}m \pgfmathprintnumber[fixed,precision=0]{\seconds}s"}{"\pgfmathprintnumber{\seconds}s"}\pgfmathresult}}

%% file: intro.tex
\section{Introduction}

Many real-world networks naturally inherit a hierarchical structure.
Often times, these interactions among the entities in the network, provide the
means of finding underlying hierarchical organization of the network.
Examples include ranking teams in a football league~\cite{Neumann2018RankingTT},
exploring  rankings in social
networks~\cite{maiya2009inferring,gupte2011finding}, terrorist
networks~\cite{memon2008retracted}, and animal
networks~\cite{jameson1999finding}.

The problem of ranking an individual,  based on its interactions with others,
has drawn attention over past
decades~\cite{gupte2011finding,nikolaj2017tiers,Neumann2018RankingTT}.
More formally, given a graph $G$ we are looking to assign each node $i$, a rank $r(i)$, an integer that minimizes a penalty.
Here the score, called \emph{agony},  penalizes the backward edges $(u, v)$ based on the difference between the ranks $r(u) - r(v)$.

Many practical networks contain information beyond the network
structure. Such labels can be used in
numerous applications such as community
detection and clustering~\cite{pool2014description,
bothorel2015clustering,galbrun2014overlapping,bai2020towards,falih2018community}.
Here, the labels are used to explain the obtained results,
which ideally increase the usability of these results to the practitioners.

In this paper, we propose an approach to infer \emph{explainable} hierarchies in labeled, weighted, directed  networks.
More specifically, we look for a decision tree-like structure, which we refer as label tree. The tree uses labels
to rank the nodes. As a quality of measure we use the agony score.

We show that our problem is \np-hard, and inapproximable even if we limit the
number of leaves in the label tree. This is in contrast with the original
optimization problem: a ranking minimizing agony can be discovered in
polynomial time.

Due to the \np-hardness of our problem, we resort to heuristics
and propose a divide-and-conquer heuristic algorithm in Section~\ref{sec:algorithm} that runs in
$\bigO{(n + m) \log n + \ell R}$ time, where $R = \sum_v \abs{L(v)}$
is the number of node-label pairs in the given graph, $\ell$ is the number of
nodes in the resulting label tree, and $n$ and  $m$ denote the number of nodes
and edges respectively.

The algorithm starts with an empty tree and finds the best split minimizing the
score. This process is continued until we have reached the maximum number of
nodes or the score can no longer be increased. It is important to point out
that the bottleneck here is computing the score while looking for the optimal
split as a naive implementation would require to enumerate over all edges for
every possible candidate.  We avoid unnecessary  enumerating over the edges by
repurposing some results from~\citet{nikolaj2017tiers} and maintaining certain
counters.  Such bookkeeping allows a fast, practical method to find label trees
with low agony score.

Furthermore, we show experimentally in Section~\ref{sec:exp} that the algorithm performs well in
practice, finds the hidden hierarchical structure in synthetic data, and
finds explainable hierarchies in real-world datasets.
Moreover, our algorithm is reasonably fast in practice as we are able to
process large networks with over hundreds of thousands of edges in less than 30 minutes.

%% file: prel.tex
\section{Preliminary notation and problem definition}\label{sec:prel}

The main input to our problem is a \emph{weighted, directed, and labeled graph} which we 
denote by $G = (V, E, w, L)$, where $V$ is a set of nodes, $E$ is a set of edges, $w$ is a function mapping an edge to a real
positive number. If $w$ is not provided, we assume that each edge has a weight
of 1. $L$ is a \emph{label} function which assigns a set of labels from $U$ to each vertex, where $U$ is the label universe. Note that, each node can have multiple labels. We  often denote $n = \abs{V}$ and $m = \abs{E}$. Given a subset of nodes $W$, we will write $E(W)$ to be the edges
having both endpoints in $W$. Given two subsets of nodes $W_1$ and $W_2$, we write $E(W_1, W_2)$
to be the edges that go from $W_1$ to $W_2$.

Given $G= (V, E, w, L)$, our goal is to rank a set of vertices $V$.  We
express this ranking with a \emph{rank assignment} $r$; a function
mapping a vertex to an integer. 

Given a graph $G$ and a rank assignment $r$, we  say that an edge
$(u, v)$ is \emph{forward} if $r(u) < r(v)$, otherwise edge is
\emph{backward}, even if $r(u) = r(v)$.
Ideally, rank assignment $r$ should not have backward edges, meaning that, for any
$(u, v) \in E$ we should have $r(u) < r(v)$. 

Next, we define the
penalty function  $\pen{}$ which penalizes the backward edges based on the severity of the violation. The penalty
for a single edge $(u, v)$ is  equal to $\pen{d} = \max(0, d + 1)$, where $d = r(u) - r(v)$. The forward edges will always receive $0$ penalty. Whenever $d \geq 0$, then the backward edges will receive $d+1$ amount of penalty. 
For example, an edge $(u, v)$ with $r(u) = r(v)$ has a penalty score of $1$. if $r(u) = r(v) + 1$, then the penalty becomes  $2$, and so on.

Next, we define a score for the ranking, to be the  sum of all backward edge penalties multiplied by its weight as follows.
\begin{definition}
Assume a weighted directed graph $G = (V, E, w)$, a rank assignment $r$, and a penalty function $\pen{}$.
We define \emph{agony} score $\score{G, r}$ as 
\[
	\score{G, r} = \sum_{e = (u, v) \in E} w(e) \pen{r(u) - r(v)}\quad.
\]
\end{definition}

Finding rank assignment minimizing agony can be done in polynomial time~\citep{gupte2011finding,nikolaj2017tiers}.

As mentioned earlier we are interested in explainable rankings. Let us now formalize this notion.

\begin{definition}
We define a \emph{label tree}, $T$ as follows. 
Label tree is an ordered, binary tree, where each leaf node represents a rank, which is an index starting from the leftmost leaf to the rightmost leaf. 
Each of the non-leaf node has a label $t$
and a Boolean value $c$.
\end{definition}

We use the label tree $T$ to partition the vertices of $G$ by traversing them
from the root to a leaf. Let $\alpha$ be a non-leaf node in $T$ with a label $t$ and a boolean value $c$.
If $c$ is true and, then a node $v$ with $t \in L(v)$ traverses to the left branch,
otherwise to the right. If $c$ is false, the branch roles are reversed.

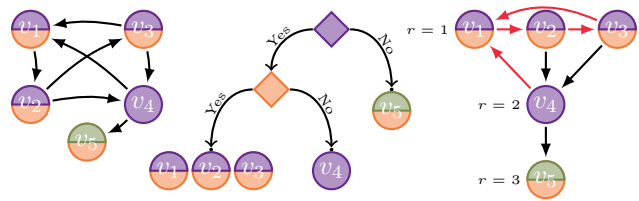
\begin{figure}
\begin{tikzpicture}

\draw[thick,fill=yafcolor2!50!white, draw=yafcolor2] (0,0) circle (7pt);
\draw[thick,fill=yafcolor1!50!white, draw=yafcolor1] (0,0) --  +(0:7pt) arc (0:180:7pt) -- cycle;
\node[text=white, circle, inner sep=2pt] (v1) at (0, 0) {$v_1$};

\draw[thick,fill=yafcolor2!50!white, draw=yafcolor2] (0,-1) circle (7pt);
\draw[thick,fill=yafcolor1!50!white, draw=yafcolor1] (0,-1) --  +(0:7pt) arc (0:180:7pt) -- cycle;
\node[text=white, circle, inner sep=2pt] (v2) at (0, -1) {$v_2$};

\draw[thick,fill=yafcolor2!50!white, draw=yafcolor2] (1.5,0) circle (7pt);
\draw[thick,fill=yafcolor1!50!white, draw=yafcolor1] (1.5,0) --  +(0:7pt) arc (0:180:7pt) -- cycle;
\node[text=white, circle, inner sep=2pt] (v3) at (1.5, 0) {$v_3$};

\draw[thick,fill=yafcolor1!50!white, draw=yafcolor1] (1.5,-1) circle (7pt);
\node[text=white, circle, inner sep=2pt] (v4) at (1.5, -1) {$v_4$};

\draw[thick,fill=yafcolor2!50!white, draw=yafcolor2] (0.75,-1.5) circle (7pt);
\draw[thick,fill=yafcolor3!50!white, draw=yafcolor3] (0.75,-1.5) --  +(0:7pt) arc (0:180:7pt) -- cycle;
\node[text=white, circle, inner sep=2pt] (v5) at (0.75, -1.5) {$v_5$};

\draw[thick, ->, >=latex] (v1) edge[bend left=10] (v2);
\draw[thick, <-, >=latex] (v1) edge[bend left=10] (v3);
\draw[thick, ->, >=latex] (v2) edge[bend left=10] (v3);
\draw[thick, ->, >=latex] (v2) edge[bend left=10] (v4);
\draw[thick, ->, >=latex] (v3) edge[bend left=10] (v4);
\draw[thick, ->, >=latex] (v4) edge[bend left=10] (v5);
\draw[thick, ->, >=latex] (v4) edge[bend right=10] (v1);

\begin{scope}[xshift=4cm]
\node[diamond, thick,fill=yafcolor1!50!white, draw=yafcolor1] (d1) at (0,0) {};
\node[diamond, thick,fill=yafcolor2!50!white, draw=yafcolor2] (d2) at (-0.8,-0.8) {};
\node[circle, fill, inner sep=0.5pt] (d3) at (-1.6,-1.6) {};
\node[circle, fill, inner sep=0.5pt] (d4) at (0,-1.6) {};
\node[circle, fill, inner sep=0.5pt] (d5) at (0.8,-0.8) {};

\draw[thick, ->] (d1) edge[out=180, in=90] node[pos=0.5, font=\tiny, sloped, auto, inner sep=0.5pt] {Yes} (d2);
\draw[thick, ->] (d2) edge[out=180, in=90] node[pos=0.5, font=\tiny, sloped, auto, inner sep=0.5pt] {Yes}  (d3);
\draw[thick, ->] (d2) edge[out=0, in=90] node[pos=0.5, font=\tiny, sloped, auto, inner sep=0.5pt] {No}  (d4);
\draw[thick, ->] (d1) edge[out=0, in=90] node[pos=0.5, font=\tiny, sloped, auto, inner sep=0.5pt] {No}  (d5);

\draw[thick,fill=yafcolor2!50!white, draw=yafcolor2] (d3) ++(-16pt, -8pt)  circle (7pt);
\draw[thick,fill=yafcolor1!50!white, draw=yafcolor1] (d3) ++(-16pt, -8pt) --  +(0:7pt) arc (0:180:7pt) -- cycle;
\draw (d3) ++(-16pt, -8pt) node[text=white, circle, inner sep=2pt] {$v_1$};

\draw[thick,fill=yafcolor2!50!white, draw=yafcolor2] (d3) ++(0, -8pt)  circle (7pt);
\draw[thick,fill=yafcolor1!50!white, draw=yafcolor1] (d3) ++(0, -8pt) --  +(0:7pt) arc (0:180:7pt) -- cycle;
\draw (d3) ++(0, -8pt) node[text=white, circle, inner sep=2pt] {$v_2$};

\draw[thick,fill=yafcolor2!50!white, draw=yafcolor2] (d3) ++(16pt, -8pt)  circle (7pt);
\draw[thick,fill=yafcolor1!50!white, draw=yafcolor1] (d3) ++(16pt, -8pt) --  +(0:7pt) arc (0:180:7pt) -- cycle;
\draw (d3) ++(16pt, -8pt) node[text=white, circle, inner sep=2pt] {$v_3$};

\draw[thick,fill=yafcolor1!50!white, draw=yafcolor1] (d4) ++(0, -8pt) circle (7pt);
\draw (d4) ++(0, -8pt) node[text=white, circle, inner sep=2pt] {$v_4$};

\draw[thick,fill=yafcolor2!50!white, draw=yafcolor2] (d5) ++(0, -8pt)  circle (7pt);
\draw[thick,fill=yafcolor3!50!white, draw=yafcolor3] (d5) ++(0, -8pt) --  +(0:7pt) arc (0:180:7pt) -- cycle;
\draw (d5) ++(0, -8pt) node[text=white, circle, inner sep=2pt] {$v_5$};

\end{scope}

\begin{scope}[xshift=6.1cm]
\draw[thick,fill=yafcolor2!50!white, draw=yafcolor2] (-0.2,0) circle (7pt);
\draw[thick,fill=yafcolor1!50!white, draw=yafcolor1] (-0.2,0) --  +(0:7pt) arc (0:180:7pt) -- cycle;
\node[text=white, circle, inner sep=2pt] (v1) at (-0.2, 0) {$v_1$};

\draw[thick,fill=yafcolor2!50!white, draw=yafcolor2] (0.75,0) circle (7pt);
\draw[thick,fill=yafcolor1!50!white, draw=yafcolor1] (0.75,0) --  +(0:7pt) arc (0:180:7pt) -- cycle;
\node[text=white, circle, inner sep=2pt] (v2) at (0.75, 0) {$v_2$};

\draw[thick,fill=yafcolor2!50!white, draw=yafcolor2] (1.7,0) circle (7pt);
\draw[thick,fill=yafcolor1!50!white, draw=yafcolor1] (1.7,0) --  +(0:7pt) arc (0:180:7pt) -- cycle;
\node[text=white, circle, inner sep=2pt] (v3) at (1.7, 0) {$v_3$};

\draw[thick,fill=yafcolor1!50!white, draw=yafcolor1] (0.75,-1) circle (7pt);
\node[text=white, circle, inner sep=2pt] (v4) at (0.75, -1) {$v_4$};

\draw[thick,fill=yafcolor2!50!white, draw=yafcolor2] (0.75,-2) circle (7pt);
\draw[thick,fill=yafcolor3!50!white, draw=yafcolor3] (0.75,-2) --  +(0:7pt) arc (0:180:7pt) -- cycle;
\node[text=white, circle, inner sep=2pt] (v5) at (0.75, -2) {$v_5$};

\node[anchor=east, font=\tiny, inner sep=1pt] at (v1.west) {$r = 1$};
\node[anchor=east, font=\tiny, inner sep=1pt] at (v4.west) {$r = 2$};
\node[anchor=east, font=\tiny, inner sep=1pt] at (v5.west) {$r = 3$};

\draw[thick, ->, >=latex, yafcolor4] (v1) edge[bend left=0] (v2);
\draw[thick, <-, >=latex, yafcolor4] (v1) edge[bend left=24] (v3);
\draw[thick, ->, >=latex, yafcolor4] (v2) edge[bend left=0] (v3);
\draw[thick, ->, >=latex] (v2) edge[bend left=0] (v4);
\draw[thick, ->, >=latex] (v3) edge[bend left=0] (v4);
\draw[thick, ->, >=latex] (v4) edge[bend left=0] (v5);
\draw[thick, ->, >=latex, yafcolor4] (v4) edge[bend right=0] (v1);
\end{scope}

\end{tikzpicture}
\caption{Graph $G$, label tree $T$, and the resulting hierarchy. The agony $\labelscore{G, T} =5$.}
\label{fig:toy}
\end{figure}

The following notations will be useful in the next section. Given a graph $G$,
a label tree $T$, and a node $\alpha$ in $T$, we write $V(\alpha)$ to be the nodes in $G$
that traverse through or end up in $\alpha$. We also write $E(\alpha) = E(V(\alpha))$.
Given a label $t$, we write $V(\alpha, t)$ to be the nodes in $V(\alpha)$ that have label $t$.
Finally, given a label $t$ and Boolean value $c$, we write $V(\alpha, t, c)$ to
be $V(\alpha, t)$ if $c$ is true, and $V(\alpha) \setminus V(\alpha, t)$ if $c$ is false.

Next, we define the score for the label tree.

\begin{definition}
Let $r(T, X)$ denote the index of the leaf after  traversing the label tree $T$ using $X$, where  $X$ is the label set. We define the \emph{agony score} $\labelscore{G, T}$ as
\[
	\labelscore{G, T} = \sum_{e = (u, v) \in E} w(e) \pen{r(T, L(u)) - r(T, L(v))}\quad.
\]
\end{definition}

Finally, we state our main optimization problem.

\begin{problem}[\prblagy]
Given a weighted, directed, labeled graph  $G = (V, E, w, L)$ and an integer $k$, 
find a label tree $T$, with at most $k$ number of leaves, which minimizes the label agony $\labelscore{G, T}$.
\end{problem}

An example of a graph and a tree is given in Fig.~\ref{fig:toy}.

We should point out that the cardinality constraint $k$ is an optional parameter,
the problem makes sense even if we set $k = \infty$.

%% file: algorithm.tex
\section{Algorithm for finding tree}\label{sec:algorithm}

In this section we first show that our problem is \np-hard and
present our greedy method for finding trees.

\subsection{Computational complexity of \prblagy}

Finding rank $r$ minimizing $\score{G, r}$ can be done
in polynomial time. However, our problem, that is, finding label tree
minimizing $\labelscore{G, T}$ turns out to be \np-hard.

\begin{proposition} 
\label{prop:np}
\prblagy is \np-hard.
\end{proposition} 

See Appendix (in the extended version of the paper) for all the proofs.

The proof of Proposition~\ref{prop:np} shows
that deciding whether there is a tree $T$ yielding
$\labelscore{G, T} = 0$ is \np-complete. This immediately implies that
there is no approximation algorithm for \prblagy since any approximation
algorithm should be able to find optimal tree $T$ if there is a solution
yielding $\labelscore{G, T} = 0$.

\begin{corollary}
\prblagy is inapproximable, unless $\poly=\np$.
\end{corollary}

The proof of
Proposition~\ref{prop:np} relied that we set cardinality constraint $k$.
It turns out that the problem is also \np-hard even if the constraint
is not enforced.

\begin{proposition} 
\label{prop:np2}
\prblagy is \np-hard, even with $k = \infty$.
\end{proposition}

\subsection{Greedy algorithm}

Since we cannot solve \prblagy efficiently  or even approximate the constrained version of our
problem, we need to resort to heuristics.

We approach the problem with a greedy method: starting from an
empty tree, we find the best possible split, perform the split
if there is a gain in score, and recurse on both leaves.
The pseudo-code for the algorithm is given in Algorithm~\ref{alg:parttion-fast}.

Here, we ignore any cardinality constraint. We will address the constraint in Section~\ref{sec:card}.

\algpartition uses two subroutines: \algsplit to compute the gain
of split candidate, and \algleft to update the tree.
In the next two sections, we discuss how these two subroutines
can be implemented efficiently.

\begin{algorithm}[ht!]
\caption{$\algpartition(\alpha)$. Calls \algsplit to find the best split. Calls \algleft to update the structures. Recurses to \algpartition for further splits.}
\label{alg:parttion-fast}

find optimal label $t$ and criterion $c$ using \algsplit\;
$\Delta \define $ reduction in score when splitting with $(t, c)$\;

\If {$\Delta < 0$} {
	
	${\beta,\gamma} \define \algleft(\alpha, t, c)$\;
	$\gain{\alpha} \define \Delta $\;
	
	$\algpartition(\beta)$; $\algpartition(\gamma)$\;
   
}

\end{algorithm}

\subsection{Computing gain}

The computational bottleneck of the greedy algorithm is finding the next split:
Given a rank function, computing agony from scratch requires $\bigO{m}$ time.
Moreover, this calculation needs to be done for every viable label.

Luckily, we can speed this process by adopting the ideas presented
by~\citet{nikolaj2017tiers}, where the authors propose a divide-and-conquer algorithm
for finding ranks with low agony $\score{G, r}$. Since the original algorithm does not
use any label information, we will modify the approach so that it can be used for our problem.

More formally, we will maintain several counters that allow us to compute
agony without enumerating over the edges.

Let $T$ be a current tree and let $\alpha$ be a leaf in $T$ with a rank $i$.
Let $U$ be the vertices with smaller rank, that is,
\[
	U = \set{v \in U \mid r(T, L(v)) < i}\quad.
\]
Similarly, let $W$ be the vertices with the larger rank
\[
	W = \set{v \in U \mid r(T, L(v)) > i}\quad.
\]

We maintain 3 counters for each leaf and each vertex: Firstly, we maintain the total weight of backward
edges that go over $\alpha$,
\begin{equation}
	\back{\alpha} = \sum_{e \in E(W, U)} w(e),
	\label{eq:count1}
\end{equation}
secondly we maintain the total weight of backward edges from higher ranks to $V(\alpha)$,
\begin{equation}
	\inback{\alpha} = \sum_{\mathclap{v \in V(\alpha)}}\inback{v}, \quad \text{where}\quad  \inback{v} = \sum_{\mathclap{e \in E(W, v)}} w(e),
	\label{eq:count2}
\end{equation}
and we maintain the total weight of backward edges from $V(\alpha)$ to lower ranks,
\begin{equation}
	\outback{\alpha} = \sum_{\mathclap{v \in V(\alpha)}}\outback{v}, \quad \text{where}\quad  \outback{v} = \sum_{\mathclap{e \in E(v, U)}} w(e)\ .
	\label{eq:count3}
\end{equation}

Finally, for each vertex $v$ in $\alpha$ we maintain the total weight of backward incoming edges
minus the total weight of backward outgoing edges,
\begin{equation}
	\diff{v} = \sum_{e \in E(W \cup V(\alpha), v)} w(e) - \sum_{e \in E(v, U \cup V(\alpha))} w(e)\ .
	\label{eq:count4}
\end{equation}

We will use the following result to compute the gain.

\begin{proposition}[Proposition~8~in~\citep{nikolaj2017tiers}]
\label{prop:split}
Let $\alpha$ be a leaf of a tree $T$. 
Assume a new tree $T'$, where we have split $\alpha$ to two leaves. Let $Y_1$ be the
vertex set of the new left leaf, and $Y_2$ the vertex set of the new right leaf. Then
the score difference is
\begin{equation}
\label{eq:gain1}
	\score{T'} - \score{T} = \back{\alpha} + \inback{\alpha} - \sum_{y \in Y_2} \diff{y} 
\end{equation}
that can be rewritten as
\begin{equation}
\label{eq:gain2}
	\score{T'} - \score{T} = \back{\alpha} + \outback{\alpha} + \sum_{y \in Y_1} \diff{y}\quad. 
\end{equation}
\end{proposition}

We make two observations:
(1) We do not need to enumerate over edges.
(2) We can test both cases for split $(t, c)$ using only $\labeldic{\alpha, t}$:
if $c = \mathit{true}$,
then we set $Y_1 = \labeldic{\alpha, t}$ and use Eq.~\ref{eq:gain2},
if $c = \mathit{false}$,
then we set $Y_2 = \labeldic{\alpha, t}$ and use Eq.~\ref{eq:gain1}.
The pseudo-code for this procedure is given in Algorithm~\ref{alg:split}.

\begin{algorithm}[ht!]
\caption{$\algsplit(\alpha,t)$, computes the gain difference of $\alpha$ due to
split based on label $t$ and criterion $c$. Tests both criteria $c = \mathit{true}$
and $c = \mathit{false}$ and returns the better.
}
\label{alg:split}
	$\Delta_1 \define \back{\alpha} + \outback{\alpha}  + \sum_{y \in \labeldic{\alpha,t}} \diff{y; \alpha}$\; 
	$\Delta_2 \define \back{\alpha} + \inback{\alpha} - \sum_{y \in \labeldic{\alpha,t}} \diff{y; \alpha}$\; 
	\Return $\min (\Delta_1, \Delta_2)$, $\Delta_1 \leq \Delta_2$\;
\end{algorithm}

\subsection{Maintaining the tree}

Once we have found the best candidate for splitting $\alpha$, we need to split
the leaf $\alpha$ into new leaves $\beta$, and $\gamma$.

More specifically, we need to update the counters $\back{\cdot}$,
$\inback{\cdot}$, $\outback{\cdot}$, and $\diff{\cdot}$, as well as compute
$V(\beta)$ and $V(\gamma)$. Here, we modify the algorithm in~\citep{nikolaj2017tiers} to suit
our needs.

Computing the vertex sets $V(\beta)$ and $V(\gamma)$ can be done
in $\bigO{\abs{V(\alpha, t)}}$ time by first moving $V(\alpha)$
to one of the child, say $\gamma$, in constant time,
and then moving the extra nodes in $\bigO{\abs{V(\alpha, t)}}$ time.

In order to update the counters we will need to iterate over the cross-edges between
$\beta$ and $\gamma$. We can do this iteration either by going over the edges
adjacent to $V(\beta)$, or by going over the edges adjacent to $V(\gamma)$. 

Both approaches will yield the same result, so we use the one that
visits fewer unnecessary edges. More specifically: if $\abs{V(\beta)} + \abs{E(\beta)} \leq \abs{V(\gamma)} + \abs{E(\gamma)}$,
enumerate over $V(\beta)$, otherwise we enumerate over $V(\gamma)$.

The pseudo-code for \algleft is given in Algorithm~\ref{alg:left}.
A straightforward calculation, which we will omit, starting from Eqs.~\ref{eq:count1}--\ref{eq:count4}
demonstrates that Algorithm~\ref{alg:left} maintains the counters correctly.
Moreover, since we need the cross-edges between $\beta$ and $\gamma$ only
once, we will delete them as we are processing them.

\begin{algorithm}[ht!]
\caption{$\algleft(\alpha)$, splits $\alpha$}
\label{alg:left}
	$N \define \labeldic{\alpha, t, c}$\;
	$P \define V(\alpha) \setminus N$\;
	
	create a new leaf $\beta$ with  $V(\beta) = N$\; 

	create a new leaf $\gamma$ with  $V(\gamma) = P$\;

	\uIf {$\abs{V(\beta)} + \abs{E(\beta)} \leq \abs{V(\gamma)} + \abs{E(\gamma)}$} {

		$\inback{\beta} \define \sum_{x \in N} \inback{x}$;
		$\inback{\gamma} \define \inback{\alpha} - \inback{\beta}$\;
		$\outback{\beta} \define \sum_{x \in N} \outback{x}$;
		$\outback{\gamma} \define \outback{\alpha} - \outback{\beta}$\;
		$\back{\beta} \define \back{\alpha} + \outback{\gamma}$;
		$\back{\gamma} \define \back{\alpha} + \inback{\beta}$\;

		\ForEach {$x \in N$} {
			\ForEach {$e = (z, x)$ such that $z \in P$} {
				add $w(e)$ to $\inback{x}$, $\inback{\beta}$, $\outback{z}$, $\outback{\gamma}$\;
				delete $e$\;
			}
			\ForEach {$e = (x, z)$ such that $z \in P$} {
				decrease $\diff{x}$, $\diff{z}$ by $w(e)$\;
				delete $e$\;
			}
		}
	}
	\Else {
		$\inback{\gamma} \define \sum_{x \in P} \inback{x}$;
		$\inback{\beta} \define \inback{\alpha} - \inback{\gamma}$\;
		$\outback{\gamma} \define \sum_{x \in P} \outback{x}$;
		$\outback{\beta} \define \outback{\alpha} - \outback{\gamma}$\;
		$\back{\beta} \define \back{\alpha} + \outback{\gamma}$;
		$\back{\gamma} \define \back{\alpha} + \inback{\beta}$\;

		\ForEach {$z \in P$} {
			\ForEach {$e = (z, x)$ such that $x \in N$} {
				add $w(e)$ to $\inback{x}$, $\inback{\beta}$, $\outback{z}$, $\outback{\gamma}$\;
				delete $e$\;
			}
			\ForEach {$e = (x, z)$ such that $x \in N$} {
				decrease $\diff{x}$, $\diff{z}$ by $w(e)$\;
				delete $e$\;
			}
		}
	}
    \Return $\beta,\gamma$ \;
\end{algorithm}

Next we prove the computational complexity of \algpartition.

\begin{proposition}
\label{prop:time}
The running time of
\algpartition is in $\bigO{(n + m) \log n + \ell R}$,
where $R = \sum_v \abs{L(v)}$ is the number of node-label pairs in $G$,
and $\ell$ is the number of nodes in the resulting label tree.
\end{proposition}

\subsection{Applying cardinality constraint by pruning}
\label{sec:card}

As our last step, we look on how to enforce possible cardinality constraint.
Given the cardinality constraint $k$ and a label tree $T$ produced by \algpartition, we can reduce the
number of leaves by pruning some branches of $T$.  The optimal subtree can be
obtained using dynamic programming, as proposed by~\citet{nikolaj2017tiers}.

Let us define $\opt{\alpha; h}$ be the optimal gain that obtained in the branch in $T$ starting from the node $\alpha$
by using only $h$ leaves (and pruning the remaining nodes).

If $\alpha$ is the root of $T$, then $\opt{\alpha; k}$
is the optimal agony achieved by pruning $T$ to have only $k$ leaves.
We initialize the array by setting  $\opt{\alpha; 1} = 0$ for any $\alpha$, and $\opt{\alpha; h} = 0$ if $\alpha$ is a leaf in $T$. If $\alpha$ is a non-leaf and $k > 1$, then we use the identity.
\[
	\opt{\alpha; h} = \gain{\alpha} + \min_{1 \leq \ell \leq h - 1} \opt{\beta; \ell} + \opt{\gamma; h - \ell},
\]
where $\beta$ is the left child and $\gamma$ is the right child, and $\gain{\alpha}$ is the improvement in agony as recorded in \algpartition.
To perform the trace-back, we further record the optimal index $\ell$ in a different table. Computing  $\opt{\alpha; h}$ requires $\bigO{k}$ time. We compute
at most $\bigO{nk}$ entries which leads to $\bigO{nk^2}$ running time.

%% file: related.tex
\section{Related work}\label{sec:related}

The notion of agony score for  ranking was first proposed by \citet{gupte2011finding} for
the unweighted, directed graphs. The authors provided a
polynomial-time, exact algorithm in which agony minimization problem has been
formulated as an integer linear program. They exploits primal-dual techniques
where the primal problem tries to minimize the agony and  the dual  finds the
maximum eulerian subgraph. Later, Tatti extended agony minimization problem
to handle weights and the cardinality constraint~\cite{nikolaj2017tiers}. At the same time, the author  introduced a
greedy heuristic  to find hierarchies provably  faster.
We should stress that these algorithms do not use any label information whereas
we propose an optimization problem and a heuristic algorithm in which the node
labels are exploited.

The appeal of using agony is that it can be solved in polynomial time.
Alternatively, if we use
constant penalty function for backward edges, the minimization problem is equivalent to \textsc{feedback arc set}
(\fasprb), which is known to be \apx-hard with a coefficient of $c =
1.3606$~\cite{dinur2005hardness}, no known constant-ratio
approximation algorithm with the best known approximation algorithm
yielding a ratio $\bigO{\log n \log \log n}$~\cite{even1998approximating}. Interestingly, while
Tatti~\cite{nikolaj2017tiers} showed
minimizing agony for a convex penalty, that is, $\score{G, r}$ can be done
in polynomial time; however, in this paper, we show that finding label tree
minimizing $\labelscore{G, T}$ turns out to be \np-hard.

Let us now look at alternative methods for ranking nodes in a graph.
We should stress that these methods do not use any label information.

A popular and a classic method for ranking players in competitions, is Elo
ranking, originally proposed as a rating method to rank chess players, based on
the outcomes of tournament games~\cite{elo1978rating}. 

\citet{maiya2009inferring} inferred  rankings, by first inducing interaction
model to the network and then searching maximum likelihood hierarchy  for each
candidate model, with a greedy heuristic.

Another way of finding hierarchies is to use graph theoretic centrality
measures. The notion of  eigenvector centrality score and degree has been
proposed by \citet{memon2008retracted}, assuming that high scores imply  highly
ranked entities. Utilizing a weighted combination of such graph theoretic
measures, including  number of cliques and betweenness  centrality was proposed
by~\citet{rowe2007automated}.
Highly ranked nodes in directed graphs typically have many outgoing edges.
Finding such nodes can be also done using a classic HITS method introduced
by~\citet{kleinberg1999authoritative}, where the score of a hub is based on the quality
of the nodes it points.

%% file: exp.tex
\section{Experimental evaluation}\label{sec:exp}

In this section we present our experimental evaluation which has the following goals: 
($i$)  Test how well \algpartition finds ground-truth hierarchy in synthetic dataset,
  and assess the impact of prevailing  noise level in labels, in terms of the Kendall's $\tau$ coefficient.\!\footnote{A measure of rank correlation which indicates the similarity between two rankings.}
($ii$) Compare the agony scores of \algpartition against the scores of hierarchies obtained without using any labels.
	Here we used \cite{nikolaj2017tiers}, which we call \algexact.
($iii$) Study how the constraint $k$ influences the agony score.
($iv$) Explore the hierarchies in real-world labeled datasets.

We implemented the algorithm in Python\footnote{See \url{https://version.helsinki.fi/dacs/} for the code.
\label{foot:code}}
and used a 2.4GHz Intel Core i5 processor and 16GB RAM.

\paragraph{Synthetic datasets:}

To test our algorithm, we generated  $7$ synthetic networks in which the characteristics are stated in
Table~\ref{tab:stats1}.
For each network, we first set our basic parameters; number of nodes $n$, number of edges $m$, and number of ranks $h$. Next we  randomly assigned the 
ranks for each node, with equal probability.
Note that, each rank has its own specific label, we call them as \emph{true} labels. Initially,  we  assigned true labels to  vertices such that  a node which belongs to the rank $r_{i}$, receives  the label $l_{i}$. This can be considered as true label assignment which is done prior to the noisy label assignment. Afterwards, we fixed two  noise parameters, $\theta$ and $\mu$, where $\theta$ indicates the percentage of nodes which contains \emph{false}  labels and $\mu$ indicates the percentage of nodes which contains \emph{noise}  labels.  To add those \emph{false} labels, we first chose a subset of vertices in accordance with $\theta$ and then added a random label chosen from  label universe $U_{true}$ to each node, without repeating the available labels for that node. Similarly, we assigned  \emph{noise} labels for a percentage of nodes determined by parameter $\mu$, randomly chosen from  $U_{noise}$, a label set of size $100$. 

After being set all noise parameter values, we fixed a certain percentage value, $\eta$ as a controlling parameter of the proportion of forward edges. 
Next, we uniformly sampled $m$ number of vertex pairs as follows:
For each rank $i = 1, \ldots, h - 1$, we generated $m / (h - 1)$ edges, between the nodes with rank $i$ and rank $i + 1$.
For each vertex pair, we determined whether the chosen edge is either forward or backward. 
This choice can be viewed in terms of a
coin-flipping experiment of a biased coin with probability $\eta$.
Finally, we removed all self-loops and aggregated the same directed edges together by assigning their accumulated  weights.

\input{statstable}

\input{statstable-com}

\paragraph{Real-world datasets:}

We explore label hierarchies in $7$ publicly available, labeled, real-world datasets. The details of the  datasets  are shown in Table~\ref{tab:stats3}. \dtname{EIES} is a  network of researchers working on social network analysis, known as \emph{Freeman's EIES networks}\footnote{\url{https://toreopsahl.com/datasets/}}. It contains a set of message links among $32$ researchers, weighted based on the number of messages sent. Each researcher is attributed based on his/her main disciplinary affiliation and the number of citations each researcher had in the social science citation index in $1978$.
\dtname{School} dataset\footnote{\url{http://www.sociopatterns.org/datasets/} } 
contains directed contact links between students in a high school in France, weighted by the length of interval where the contact was active. Students are labeled with their class labels. 
\dtname{Cora}~\cite{nr2015}\footnote{\url{https://networkrepository.com} \label{foot:nw-repo}}  and \dtname{Cite-seer}~\cite{nr2015}\footref{foot:nw-repo} datasets are citation networks where papers are attributed with a category and a class respectively.
\dtname{Patent-citation}~\cite{UDMS_dataset}\footnote{\url{https://www2.helsinki.fi/en/researchgroups/unified-database-management-systems-udbms/datasets/patent-dataset}}  contains patent citations data.
Each patent is a vertex and each citation is an edge in our network. Each patent has a class, a technological
category, and a technological sub-category which we consider as node labels. For our experiments, we extracted a sub-network which contains first $7\,000$ citations and then filtered out the nodes whose category details are missing from the dataset.
\dtname{DBLP-citation}~\cite{tang2008arnetminer}\footnote{\url{https://www.aminer.org/citation}} contains  citation relationships from top venues in Data Mining and Machine Learning~(SDM, NIPS, ICDM, KDD, ECMLPKDD, and WWW). We extracted set of labels by stemming the titles of the papers and removing stop words. \dtname{Physics-citation}\footnote{\url{https://github.com/chriskal96/physics-theory-citation-network}} is a citation network  extracted from Journals in the High Energy Physics Theory category. Herein, we use the title of the publication as labels, after stemming the titles and removing stop words.


\paragraph{Results of synthetic datasets:}

The detailed statistics and the running times  are given in Table~\ref{tab:stats2}.
First, we observe  $2nd$, $3rd$, and $4th$ columns in Table~\ref{tab:stats2}. These columns display  ground truth agony scores of the datasets, followed by the scores obtained with \algpartition and  \algexact.
Generally, the discovered agony score obtained by \algpartition is  more closer to the ground truth agony score, as opposed to the  score found by  \algexact.
Secondly, consider  $k\tau_{dis}$ column of Table~\ref{tab:stats2}, which exhibits the Kendall's $\tau$ measures attained by our algorithm. For all these experiments Kendall's $\tau$ coefficient  exceeds $0.9$; which indicates a high quality match of the discovered hierarchies with compared to the ground truth, even under the presence of $10\%$ false labels. 

Let us now compare Kendall's $\tau$ measures achieved by \algpartition with the scores by \algexact, which is shown in the $k\tau_{base}$ column.
We can observe that Kendall's $\tau$ measures obtained by \algpartition is significantly higher than the ones by baseline, except for one outlier case.
Next, let us consider  the $5th$ column of Table~\ref{tab:stats2}, which shows the number of ranks discovered by \algpartition. Interestingly, this is equal to the number of ground truth ranks.
Moreover, the numbers of ranks discovered by \algexact ($6th$ column) are  much higher than the ground truth.
Finally, from the
computational time results which are stated in the last column in Table~\ref{tab:stats2}, we see that the running times
are  reasonably practical to run our algorithm nearly in $3$ minutes for a graph with over $40\,000$ vertices and $100\,000$ edges.


\begin{figure}[t!]
\begin{center}
\setlength{\tabcolsep}{0pt}
\begin{tikzpicture}
\begin{axis}[xlabel={  Percentage of forward edges~($\eta$)  }, ylabel= {Kendall's  $\tau$},
    width = 6.9cm,
    height = 3.5cm,
    xmin = 20,
    xmax = 100,
    ymin = -1,
    ymax = 1,
    scaled y ticks = false,
    cycle list name=yaf,
    yticklabel style={/pgf/number format/fixed},
    legend entries = {$\theta=0\%$,$\theta=5\%$, $\theta=10\%$, $\theta=20\%$, $\theta=30\%$},
	legend pos=outer north east,
    no markers,
]
\addplot table [x=x, y=y1, col sep=comma] {agony-vs-per.csv};
\addplot table [x=x, y=y2, col sep=comma] {agony-vs-per.csv};
\addplot table [x=x, y=y3, col sep=comma] {agony-vs-per.csv};
\addplot table [x=x, y=y5, col sep=comma] {agony-vs-per.csv};
\addplot table [x=x, y=y6, col sep=comma] {agony-vs-per.csv};
\pgfplotsextra{\yafdrawaxis{20}{100}{-1}{1}}
\end{axis}
\end{tikzpicture}

\caption{Kendall's  $\tau$ coefficient  as a function of $\eta$ for the cases of several false label probabilities~($\theta$) shown in the legend. This experiment is done for  $\abs{V}=4\,000$, $\abs{E}=7\,000$, $\mu=0.05$, and $h=10$ using \algpartition.}
\label{fig:per-kendall}
\end{center}
\end{figure}
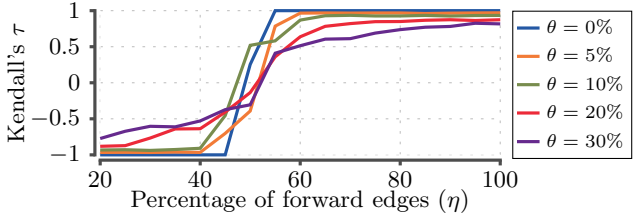


\begin{figure}[t!]
\begin{center}
\setlength{\tabcolsep}{0pt}
\begin{tikzpicture}
\begin{axis}[xlabel={Percentage of forward edges~($\eta$)  }, ylabel= {Kendall's  $\tau$},
    width = 6.9cm,
    height = 3.5 cm,
    xmin = 20,
    xmax = 100,
    ymin = -1,
    ymax = 1,
    scaled y ticks = false,
    cycle list name=yaf,
    yticklabel style={/pgf/number format/fixed},
    legend entries = {$\mu=10\%$,$\mu=20\%$, $\mu=30\%$, $\mu=50\%$, $\mu=70\%$},
	legend pos=outer north east,
    no markers,
]
\addplot table [x=x, y=y1, col sep=comma] {agony-vs-per-ind.csv};
\addplot table [x=x, y=y2, col sep=comma] {agony-vs-per-ind.csv};
\addplot table [x=x, y=y3, col sep=comma] {agony-vs-per-ind.csv};
\addplot table [x=x, y=y4, col sep=comma] {agony-vs-per-ind.csv};
\addplot table [x=x, y=y5, col sep=comma] {agony-vs-per-ind.csv};
\pgfplotsextra{\yafdrawaxis{20}{100}{-1}{1}}
\end{axis}
\end{tikzpicture}

\caption{Kendall's  $\tau$ coefficient  as a function of $\eta$  for the cases of several  noise label probabilities~($\mu$) shown in the legend. This experiment is done for  $\abs{V}=4\,000$, $\abs{E}=7\,000$, $\theta=0.15$, and $h=10$ using \algpartition.}
\label{fig:per-kendall-ind}
\end{center}
\end{figure}
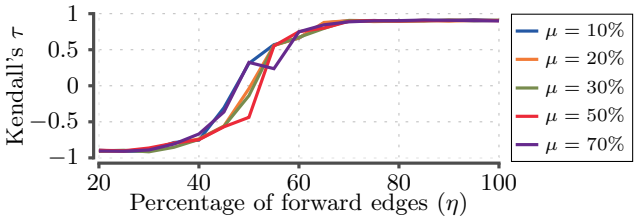

Let us now consider Figure~\ref{fig:per-kendall} which demonstrates how Kendall's $\tau$ metric changes with respect to the percentage of forward edges in the network. In general, we see that Kendall's $\tau$  gradually increases when the percentage of forward edges increases.  The coefficient  significantly increases nearly at the $50\%$ percentage, from negative to positive, due to the fact that there are more forward edges than backward edges after $50\%$. We repeat this experiment by adjusting the false label probability  $\theta$ to be $0\%, 5\%, 10\%$, $20\%$, and $30\%$. When there are no false labels,  the coefficient achieves $1$ which implies a perfect match in hierarchy, while the percentage of forward edges reaches nearly $50\%$. The coefficient  achieves its max at $60\%$ for the case of $5\%$ of false  labels, whereas same is obtained nearly at $95\%$ for the case of $30\%$ of false labels. Hence we  conclude that lesser the amount of \emph{false} label noise is present, more accurate hierarchies could be extracted and more percentage of backward edges can be withstood.

Let us now investigate the effect of noise label probability $\mu$; which is shown in Figure~\ref{fig:per-kendall-ind}. On  contrary to the previous experiment, we fix $\theta$ and then repeat the same experiment by varying   $\mu$. Unlike false label probability $\theta$, even if we increase $\mu$ to be $70\%$, we can conclude that $\mu$ does not significantly affect our performance metric,  Kendall's $\tau$ coefficient.


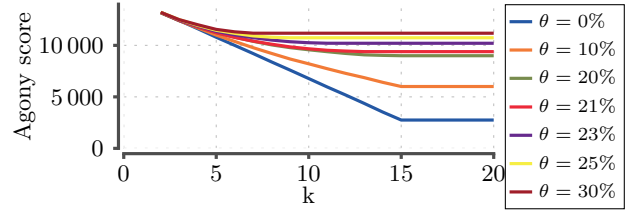
\begin{figure}[t!]
\begin{center}
\setlength{\tabcolsep}{0pt}
\begin{tabular}{ll}
\begin{tikzpicture}
\begin{axis}[xlabel={k  }, ylabel= {Agony score},
    width = 6.5cm,
    height = 3.5cm,
    xmin = 0,
    xmax = 20,
    ymin = 0,
    ymax = 14000,
    scaled y ticks = false,
    cycle list name=yaf,
    no markers,
    legend entries = { $\theta=0\%$, $\theta=10\%$, $\theta=20\%$, $\theta=21\%$, $\theta=23\%$, $\theta=25\%$, $\theta=30\%$},
	legend pos=outer north east,
    no markers,
]
\addplot table [x=x, y=y1, col sep=comma] {agony-k.csv};
\addplot table [x=x, y=y2, col sep=comma] {agony-k.csv};
\addplot table [x=x, y=y3, col sep=comma] {agony-k.csv};
\addplot table [x=x, y=y4, col sep=comma] {agony-k.csv};
\addplot table [x=x, y=y5, col sep=comma] {agony-k.csv};
\addplot table [x=x, y=y6, col sep=comma] {agony-k.csv};
\addplot table [x=x, y=y7, col sep=comma] {agony-k.csv};
\pgfplotsextra{\yafdrawaxis{0}{20}{0}{14000}}
\end{axis}
\end{tikzpicture}
\end{tabular}

\caption{Agony score  ($\score{T}$)  as a function of the constraint $k$  for the cases of
 several false label probabilities~($\theta$) as shown in the legend. This experiment is done  for  $\abs{V}=4\,000$, $\abs{E} \approx 7\,000$, $h=15$, $\mu=0.05$, and $\eta=0.9$
 using \algpartition.}
\label{fig:agony-k}
\end{center}
\end{figure}

Next Figure~\ref{fig:agony-k} demonstrates the agony score as a function of $k$.  Here we repeat the experiment at different false label probabilities such as $ 0\%, 10\%, 20\%, 21\%, 23\%, 25\%$, and $30\%$. From the results we see that, for the first $3$ cases,  in which no false labels, $10\%$  and $20\%$ of false labels, agony score decreases
more prominently until it approaches $k=15$ and then it remains constant even if we increase $k$ further. Similar phenomenon happens at $k=14$, $k=13$, $k=9$, and $k=8$, for the cases of $ 21\%, 23\%, 25\%,$ and $30\%$ of  false labels respectively. This is because in latter experiments the tree contains at most $14$, $13$, $9$,  and $8$  leaves and can not improve the agony score by splitting further. In addition, we  observe that the elbow point is reached at the earliest for \emph{high-noisy} labels, with compared to the \emph{low-noisy} labels.


\paragraph{Results of real-world datasets:}

\input{statstable-real}

\input{statstable-real-com}

For each dataset, we computed the agony using \algpartition and \algexact. The statistics of  discovered hierarchical  structures for real-world  datasets  are shown in Table~\ref{tab:stats4}. 
First, let us look at the discovered number of ranks, which are given in Table~\ref{tab:stats4}. 
Generally, $h_{base}$ is higher with compared to $h_{dis}$ except for \dtname{EIES} and \dtname{Patent-citation} datasets.
Next, $2nd$ and $3rd$ columns in Table~\ref{tab:stats4} display  agony scores obtained by \algpartition and 
\algexact. As you can see in $q_{base}$, $\score{G, r} = 0$ is reached  twice  for \dtname{Cite-seer} and \dtname{Patent-citation} datasets by \algexact. The column \emph{d} in Table~\ref{tab:stats4} shows the depth of the constructed tree which is significantly smaller than $\abs{V}$. The last column in Table~\ref{tab:stats4} shows that running times are reasonably practical so that we could construct the label tree for a label graph with more than $300\,000$ edges, $80\,000$ vertices, and $4\,000$ labels in less than $22$ minutes.


Note that we can solve \prblagy exactly
if each node has only a single label with a new graph
by using the set of labels
as new nodes and  combined edges as its edges.
We then run the \algexact algorithm for this new label-free network to obtain
the optimal rank.
We do this 
for \dtname{Cora} and \dtname{Cite-seer}
datasets as each paper has only one label. 
The results of those experiments are shown in Table~\ref{tab:stats5}.
Let us focus on  $q_{dis}$ and $q_{new}$ columns in Table~\ref{tab:stats5}. The
discovered agony scores obtained using our algorithm and baseline are
approximately similar, on newly formed label-free networks, implying that our
heuristic finds either optimal or close to optimal rankings for these networks.

\input{statstable-single-real}


\paragraph{Case study:}

To conclude the experimental section, we take a closer look at  hierarchical groups found by  \algpartition  for some of the datasets. 
We highlight two datasets, additional trees are given in Appendix.

\begin{figure}[t!]
\begin{center}
\setlength{\tabcolsep}{0pt}
\begin{tabular}{ll}

\begin{tikzpicture}[baseline = 0pt, yscale=0.5]
\tikzstyle{node1}=[inner sep=0pt]
\node[node1] (n1) at (0, 0) {$\shortstack{Theory}$};

\node[node1] (n2)  at (-2, -1.7) {$\shortstack{Braneworld}
 $};
\node[node1] (n3) at (2, -1.7) {$\shortstack{Conjecture}
 $};

\node[node1] (n4) at (-3, -3.4) {$\shortstack{Holographic}$};

\node[node1] (n10) at (-3.7, -5.0) {$ \shortstack{ Gravity}$};
\node[node1] (n14) at (-4.0, -6.4) {$ \shortstack{ \color{purple} R1}$};
\node[node1] (n15) at (-3.4, -6.4) {$ \shortstack{ \color{purple} R2}$};

\node[node1] (n11) at (-2.3, -5.0) {$ \shortstack{ Duality}$};
\node[node1] (n16) at (-2.6, -6.4) {$ \shortstack{ \color{purple} R3}$};
\node[node1] (n17) at (-2.0, -6.4) {$ \shortstack{ \color{purple} R4}$};

\node[node1] (n5) at (-1, -3.4) {$\shortstack{World}$};
\node[node1] (n12) at (-.5, -5.0) {$ \shortstack{ \color{purple} R6}$};
\node[node1] (n13) at (-1.5, -5.0) {$ \shortstack{ \color{purple} R5}$};

\node[node1] (n6) at (1, -3.4) {$\shortstack{Electric}$};
\node[node1] (n7) at (3, -3.4) {$\shortstack{ \color{purple} R9}$};

\node[node1] (n8) at (0.5, -5) {$ \shortstack{ \color{purple} R7}$};
\node[node1] (n9) at (1.5, -5) {$ \shortstack{ \color{purple} R8}$};

\draw[yafcolor3, ->, >=latex, thick, dashed, in=90, out=-90] (n1) edge (n2);
\draw[yafcolor3, ->, >=latex, thick, in=90, out=-90] (n1) edge (n3);

\draw[yafcolor3, ->, >=latex, thick, in=90, out=-90] (n2) edge (n4);
\draw[yafcolor3, ->, >=latex, thick, dashed, in=90, out=-90] (n2) edge (n5);

\draw[yafcolor3, ->, >=latex, thick, dashed, in=90, out=-90] (n3) edge (n6);
\draw[yafcolor3, ->, >=latex, thick, in=90, out=-90] (n3) edge (n7);

\draw[yafcolor3, ->, >=latex, thick, in=90, out=-90] (n4) edge (n10);
\draw[yafcolor3, ->, >=latex, thick, dashed, in=90, out=-90] (n4) edge (n11);

\draw[yafcolor3, ->, >=latex, thick, dashed, in=90, out=-90] (n5) edge (n12);
\draw[yafcolor3, ->, >=latex, thick, in=90, out=-90] (n5) edge (n13);

\draw[yafcolor3, ->, >=latex, thick, dashed, in=90, out=-90] (n6) edge (n8);
\draw[yafcolor3, ->, >=latex, thick, in=90, out=-90] (n6) edge (n9);

\draw[yafcolor3, ->, >=latex, thick, in=90, out=-90] (n10) edge (n14);
\draw[yafcolor3, ->, >=latex, thick, dashed, in=90, out=-90] (n10) edge (n15);

\draw[yafcolor3, ->, >=latex, thick, in=90, out=-90] (n11) edge (n16);
\draw[yafcolor3, ->, >=latex, thick, dashed, in=90, out=-90] (n11) edge (n17);

\end{tikzpicture}\\


\begin{tikzpicture}[baseline = 0pt, yscale=0.5,xscale=0.9]
\tikzstyle{node1}=[inner sep=0pt]

\node[node1] (n1) at (0, 0) {$\shortstack{Algorithm}$};

\node[node1] (n2)  at (-1.8, -1.7) {$\shortstack{Recommendation}$};
\node[node1] (n3) at (1.8, -1.7) {$\shortstack{Database}$};

\node[node1] (n4) at (-3.0, -3.4) {$\shortstack{Sequential}$};

\node[node1] (n10) at (-4.2, -5.0) {$ \shortstack{ Network}$};
\node[node1] (n14) at (-4.9, -6.4) {$ \shortstack{ \color{purple} R1}$};
\node[node1] (n18) at (-4.0, -7.6) {$ \shortstack{\color{purple} R2}$};
\node[node1] (n19) at (-3.0, -7.6) {$ \shortstack{ \color{purple} R3}$};

\node[node1] (n15) at (-3.5, -6.4) {$ \shortstack{ Attentional}$};

\node[node1] (n11) at (-1.8, -5.0) {$ \shortstack{ Generalized}$};
\node[node1] (n16) at (-2.3, -6.4) {$ \shortstack{ \color{purple} R4}$};
\node[node1] (n17) at (-1.3, -6.4) {$ \shortstack{ \color{purple} R5}$};

\node[node1] (n5) at (-.6, -3.4) {$\shortstack{\color{purple} R6}$};

\node[node1] (n6) at (1, -3.4) {$\shortstack{\color{purple} R7}$};
\node[node1] (n7) at (2.7, -3.4) {$\shortstack{ Discovering}$};

\node[node1] (n8) at (2.2, -5) {$ \shortstack{ \color{purple} R8}$};
\node[node1] (n9) at (3.2, -5) {$ \shortstack{ \color{purple} R9}$};

\draw[yafcolor3, ->, >=latex, thick, dashed, in=90, out=-90] (n1) edge (n2);
\draw[yafcolor3, ->, >=latex, thick, in=90, out=-90] (n1) edge (n3);

\draw[yafcolor3, ->, >=latex, thick, in=90, out=-90] (n2) edge (n4);
\draw[yafcolor3, ->, >=latex, thick, dashed, in=90, out=-90] (n2) edge (n5);

\draw[yafcolor3, ->, >=latex, thick, dashed, in=90, out=-90] (n3) edge (n6);
\draw[yafcolor3, ->, >=latex, thick, in=90, out=-90] (n3) edge (n7);

\draw[yafcolor3, ->, >=latex, thick, in=90, out=-90] (n4) edge (n10);
\draw[yafcolor3, ->, >=latex, thick, dashed, in=90, out=-90] (n4) edge (n11);

\draw[yafcolor3, ->, >=latex, thick, dashed, in=90, out=-90] (n7) edge (n8);
\draw[yafcolor3, ->, >=latex, thick, in=90, out=-90] (n7) edge (n9);

\draw[yafcolor3, ->, >=latex, thick, in=90, out=-90] (n10) edge (n14);
\draw[yafcolor3, ->, >=latex, thick, dashed, in=90, out=-90] (n10) edge (n15);

\draw[yafcolor3, ->, >=latex, thick, in=90, out=-90] (n11) edge (n16);
\draw[yafcolor3, ->, >=latex, thick, dashed, in=90, out=-90] (n11) edge (n17);

\draw[yafcolor3, ->, >=latex, thick, in=90, out=-90] (n15) edge (n18);
\draw[yafcolor3, ->, >=latex, thick, dashed, in=90, out=-90] (n15) edge (n19);

\end{tikzpicture}
\end{tabular}
\caption{Discovered label trees for \dtname{Physics} (above) and \dtname{DBLP} (below) datasets. Solid lines indicate the branch for the nodes with the corresponding label.
}
\label{fig:tree-dblp-2}
\end{center}
\end{figure}
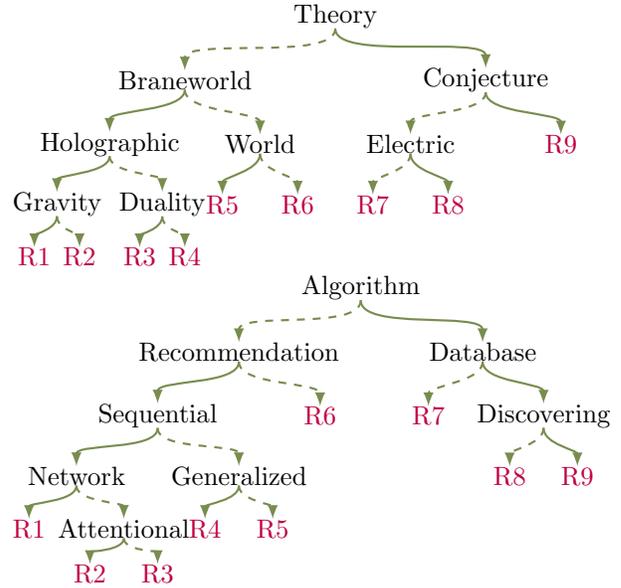

The label tree obtained for \dtname{Physics-citation} dataset is given in the top of Figure~\ref{fig:tree-dblp-2}. Our algorithm  first partitions the physics publications based on whether it is theoretical  or application-based, suggesting that application-based papers cite more theoretical papers. Next, the label \emph{conjecture} divides the set of theoretical papers further in to two halves, suggesting that  more conjectures are likely to be presented in theoretical papers. Application-based publications are then divided further with the label \emph{braneworld} which is a cosmology related scientific term. The labels like \emph{holographic}, \emph{gravity}, and \emph{braneworld} are in the same line of research  which is sensible.


The label tree obtained for \dtname{DBLP-citation} dataset is 
shown in the bottom tree of Figure~\ref{fig:tree-dblp-2}. This tree shows $9$ ranks in total. Our algorithm  first partitions the  publications based on whether the title contains the word, \emph{algorithm}, suggesting  that the papers which contain word, \emph{algorithm} are cited more by the papers which does not contain the word, \emph{algorithm}.  Likewise, this label tree postulates that the publications which contain the words,  \emph{algorithm}, \emph{database}, and  \emph{discovering} are cited more by the publications which contain \emph{recommendation}, \emph{sequential}, and  \emph{network} in their titles.

%% file: statstable.tex
\begin{table}[t!]
\setlength{\tabcolsep}{0pt}
\caption{Characteristics of the synthetic datasets. Here,  $h$ indicates the number of hierarchy levels~(ranks), $\eta$ is the percentage of forward edges, $\theta$ and $\mu$ implies the percentage of vertices which contains false labels and noise labels respectively.
}

\label{tab:stats1}
\begin{filecontents*}{table-1.csv}
name,n,m,k,dis_k,dis_k_base,P,noise,mu,agony_true,agony_found,agony_baseline,kendall,CT,depth,kendall_base
\dtname{Syn-1},4000,3200,5,5,25,0.6,0.04,0.05,2242,2322,10,0.9666105073,2.782031059,5,0.102281835
\dtname{Syn-2},5000,7000,8,8,47,0.7,0.04,0.06,4160,4453,648,0.9740631439,6.465184927,8,0.2879694756
\dtname{Syn-3},1200,1200,5,5,20,0.65,0.05,0.07,782,798,2,0.9701096689,0.4997770786,5,0.1541807669
\dtname{Syn-4},1000,1750,6,7,27,0.9,0.05,0.08,388,535,2,0.9590191811,0.6152789593,6,0.5751012947
\dtname{Syn-5},4000,4200,7,8,22,0.85,0.06,0.09,1268,1591,16,0.9592091314,4.207409859,7,0.2625473263
\dtname{Syn-6},1000,5600,15,15,19,0.8,0.08,0.1,2262,3237,2158,0.9464063246,1.294280052,15,0.9605693015
\dtname{Syn-7},40000,135000,10,10,24,0.75,0.08,0.11,66978,89017,51504,0.9326393801,189.9484668,10,0.7931704822
\end{filecontents*}
\pgfplotstabletypeset[
    begin table={\begin{tabular*}{.49\textwidth}},
    end table={\end{tabular*}},
    col sep=comma,
	columns = {name, n, m, k, noise, mu, P},
    columns/name/.style={string type, column type={@{\extracolsep{\fill}}l}, column name=\emph{Dataset}},
    columns/n/.style={fixed, set thousands separator={\,}, column type=r, column name=$\abs{V}$},
    columns/m/.style={fixed, set thousands separator={\,}, column type=r, column name=$\abs{E}$},
    columns/k/.style={fixed, set thousands separator={\,}, column type=r, column name=$h$},
    columns/P/.style={fixed, set thousands separator={\,}, column type=r, column name=$\eta$},
    columns/agony_true/.style={fixed, set thousands separator={\,}, column type=r, column name=$q_{true}$},
    columns/agony_found/.style={fixed, set thousands separator={\,}, column type=r, column name=$q_{dis}$},
    columns/mu/.style={fixed, set thousands separator={\,}, column type=r, column name=$\mu$},    
    columns/noise/.style={fixed, set thousands separator={\,}, column type=r, column name=$\theta$},   
    columns/CT/.style={fixed,set thousands separator={\,}, column type=r, column name=$t$},
    columns/baseline/.style={fixed,set thousands separator={\,}, column type=r, column name=$baseline$},   
    columns/depth/.style={fixed, set thousands separator={\,}, column type=r, column name=d},
    every head row/.style={
		before row={\toprule},
			after row=\midrule},
    every last row/.style={after row=\bottomrule},
]
{table-1.csv}
\end{table}

%% file: statstable-com.tex
\begin{table}[t!]
\setlength{\tabcolsep}{0pt}
\caption{Statistics from the experiments with the synthetic datasets. 
Here,
$q_{true}$ and $q_{dis}$ are the ground truth and discovered agony scores respectively, $q_{base}$ is the discovered agony using  \algexact, $h_{dis}$ and $h_{base}$ are the discovered number of ranks using \algpartition and \algexact, $k_{tau}$ is the Kendall's $\tau$ coefficient, and $time$ gives the computational time in seconds for \algpartition.
}

\label{tab:stats2}
\pgfplotstabletypeset[
    begin table={\begin{tabular*}{.49\textwidth}},
    end table={\end{tabular*}},
    col sep=comma,
	columns = {name,agony_true,agony_found,agony_baseline,dis_k,dis_k_base, kendall,kendall_base,CT},
    columns/name/.style={string type, column type={@{\extracolsep{\fill}}l}, column name=\emph{Dataset}},
    columns/n/.style={fixed, set thousands separator={\,}, column type=r, column name=$\abs{V}$},
    columns/m/.style={fixed, set thousands separator={\,}, column type=r, column name=$\abs{M}$},
    columns/dis_k/.style={fixed, set thousands separator={\,}, column type=r, column name=$h_{dis}$},
    columns/dis_k_base/.style={fixed, set thousands separator={\,}, column type=r, column name=$h_{base}$},
    columns/P/.style={fixed, set thousands separator={\,}, column type=r, column name=$per$},
    columns/agony_true/.style={fixed, set thousands separator={\,}, column type=r, column name=$q_{true}$},
    columns/agony_found/.style={fixed, set thousands separator={\,}, column type=r, column name=$q_{dis}$},
    columns/kendall/.style={fixed, set thousands separator={\,}, column type=r, column name=$k\tau_{dis}$},   columns/kendall_base/.style={fixed, set thousands separator={\,}, column type=r, column name=$k\tau_{base}$},  
    columns/tag/.style={string type, column type=r, column name=$tag$},    
    columns/CT/.style={fixed,set thousands separator={\,}, column type=r, column name=$time$},
    columns/agony_baseline/.style={fixed,set thousands separator={\,}, column type=r, column name=$q_{base}$},   
    columns/depth/.style={fixed, set thousands separator={\,}, column type=r, column name=$d$},
    every head row/.style={
		before row={\toprule},
			after row=\midrule},
    every last row/.style={after row=\bottomrule},
]
{table-1.csv}
\end{table}

%% file: statstable-real.tex
\begin{table}[t!]
\setlength{\tabcolsep}{0pt}
\caption{Characteristics of the real-world datasets. Here, $\abs{T}$ is the number of labels.
}

\label{tab:stats3}
\begin{filecontents*}{real-world.csv}
name,n,m,k,k_base,agony_found,agony_baseline,CT,depth,t,file,url
\dtname{EIES},32,460,4,3,14035,12682,3.4ms,4,9,,https://toreopsahl.com/datasets/#FreemansEIES
\dtname{School},329,502,3,10,1426,982,20ms,3,9,,http://www.sociopatterns.org/datasets/high-school-contact-and-friendship-networks/
\dtname{Cora},2708,5429,3,18,5351,340,35s,3,7,n-1,http://zhang18f.myweb.cs.uwindsor.ca/datasets/#
\dtname{Cite-seer},3264,4536,3,16,4397,0,1.3s,3,6,n-8,https://networkrepository.com/citeseer.php
\dtname{Patent-citation},5439,4232,7,3,4074,0,4m,4,394,real-8,https://www2.helsinki.fi/en/researchgroups/unified-database-management-systems-udbms/datasets/patent-dataset
\dtname{DBLP-citation},30581,70972,10,38,68511,329,43m,6,10245,,
\dtname{Physics-citation},29555,352807,9,236,339475,2268,22m,5,4715,,https://github.com/chriskal96/physics-theory-citation-network
\end{filecontents*}
\pgfplotstabletypeset[
    begin table={\begin{tabular*}{.49\textwidth}},
    end table={\end{tabular*}},
    col sep=comma,
	columns = {name, n, m, t},
    columns/name/.style={string type, column type={@{\extracolsep{\fill}}l}, column name=\emph{Dataset}},
    columns/n/.style={fixed, set thousands separator={\,}, column type=r, column name=$\abs{V}$},
    columns/m/.style={fixed, set thousands separator={\,}, column type=r, column name=$\abs{E}$},
    columns/k/.style={fixed, set thousands separator={\,}, column type=r, column name=$h_{dis}$},
    columns/k_base/.style={fixed, set thousands separator={\,}, column type=r, column name=$h_{base}$},
     columns/t/.style={fixed, set thousands separator={\,}, column type=r, column name=$\abs{T}$},
    columns/agony_found/.style={fixed, set thousands separator={\,}, column type=r, column name=$q_{dis}$},
    columns/CT/.style={fixed,set thousands separator={\,}, column type=r, column name=$t$},
    columns/agony_baseline/.style={fixed,set thousands separator={\,}, column type=r, column name=$q_{base}$},   
    columns/depth/.style={fixed, set thousands separator={\,}, column type=r, column name=d},
    every head row/.style={
		before row={\toprule},
			after row=\midrule},
    every last row/.style={after row=\bottomrule},
]
{real-world.csv}
\end{table}

%% file: statstable-real-com.tex
\begin{table}[t!]
\setlength{\tabcolsep}{0pt}
\caption{Experimental details with real-world datasets. 
Here, $q_{dis}$ and $q_{base}$ are discovered agony using  \algpartition and \algexact respectively,
Similarly, $h_{dis}$ and $h_{base}$ indicate the  number of ranks discovered,
$d$ gives the height of the tree constructed using \algpartition,
and $time$ gives the computational time for \algpartition.
}

\label{tab:stats4}
\pgfplotstabletypeset[
    begin table={\begin{tabular*}{.49\textwidth}},
    end table={\end{tabular*}},
    col sep=comma,
	columns = {name, agony_found,agony_baseline, k, k_base, depth,CT},
    columns/name/.style={string type, column type={@{\extracolsep{\fill}}l}, column name=\emph{Dataset}},
    columns/n/.style={fixed, set thousands separator={\,}, column type=r, column name=$\abs{V}$},
    columns/m/.style={fixed, set thousands separator={\,}, column type=r, column name=$\abs{E}$},
    columns/k/.style={fixed, set thousands separator={\,}, column type=r, column name=$h_{dis}$},
    columns/k_base/.style={fixed, set thousands separator={\,}, column type=r, column name=$h_{base}$},
     columns/t/.style={fixed, set thousands separator={\,}, column type=r, column name=$\abs{T}$},
    columns/agony_found/.style={fixed, set thousands separator={\,}, column type=r, column name=$q_{dis}$},
    columns/CT/.style={string type, column type=r, column name=$time$},
    columns/agony_baseline/.style={fixed,set thousands separator={\,}, column type=r, column name=$q_{base}$},   
    columns/depth/.style={fixed, set thousands separator={\,}, column type=r, column name=$d$},
    every head row/.style={
		before row={\toprule},
			after row=\midrule},
    every last row/.style={after row=\bottomrule},
]
{real-world.csv}
\end{table}

%% file: statstable-single-real.tex
\begin{table}[t!]
\setlength{\tabcolsep}{0pt}
\caption{Details of two label-free networks constructed using \dtname{Cora} and \dtname{Cite-seer} datasets. 
Here, $q_{dis}$ is the discovered agony using  \algpartition,  $q_{new}$ gives discovered agony using \algexact, on newly formed label-free network, $h$ indicates the discovered number of ranks using \algpartition, and $h_{new}$ indicates the discovered number of ranks using \algexact on label-free network.
}

\label{tab:stats5}
\begin{filecontents*}{single-label-real-world.csv}
name,n,m,n_new,m_new,k,k_base,k_tag_free,agony_found,agony_baseline,agony_tag_free
\dtname{Cora},2708,5429,7,39,3,18,3,5351,340,5315
\dtname{Cite-seer},3264,4536,6,30,3,16,3,4397,0,4397
\end{filecontents*}
\pgfplotstabletypeset[
    begin table={\begin{tabular*}{.49\textwidth}},
    end table={\end{tabular*}},
    col sep=comma,
	columns = {name, n_new, m_new, agony_found, agony_tag_free,k, k_tag_free},
    columns/name/.style={string type, column type={@{\extracolsep{\fill}}l}, column name=\emph{Dataset}},
    columns/n/.style={fixed, set thousands separator={\,}, column type=r, column name=$\abs{V}$},
    columns/m/.style={fixed, set thousands separator={\,}, column type=r, column name=$\abs{E}$},
    columns/n_new/.style={fixed, set thousands separator={\,}, column type=r, column name=$\abs{V_{new}}$},
    columns/m_new/.style={fixed, set thousands separator={\,}, column type=r, column name=$\abs{E_{new}}$},
    columns/k/.style={fixed, set thousands separator={\,}, column type=r, column name=$h$},
    columns/k_base/.style={fixed, set thousands separator={\,}, column type=r, column name=$h_{base}$},
    columns/k_tag_free/.style={fixed, set thousands separator={\,}, column type=r, column name=$h_{new}$},
     columns/t/.style={fixed, set thousands separator={\,}, column type=r, column name=$\abs{T}$},
    columns/agony_found/.style={fixed, set thousands separator={\,}, column type=r, column name=$q_{dis}$},
    columns/CT/.style={fixed,set thousands separator={\,}, column type=r, column name=$time$},
    columns/agony_baseline/.style={fixed,set thousands separator={\,}, column type=r, column name=$q_{base}$},
    columns/agony_tag_free/.style={fixed,set thousands separator={\,}, column type=r, column name=$q_{new}$},    
    columns/depth/.style={fixed, set thousands separator={\,}, column type=r, column name=$depth$},
    every head row/.style={
		before row={\toprule},
			after row=\midrule},
    every last row/.style={after row=\bottomrule},
]
{single-label-real-world.csv}
\end{table}

%% file: conclusions.tex
\section{Concluding remarks}\label{sec:conclusions}

We introduced a novel tree-based, hierarchy mining problem for vertex-labeled,
directed graphs. Here the goal is to rank the nodes into tiers so that
ideally the edges are directing to the lower ranks. The ranking is done with
a decision tree that can only use node labels.



The goal was to minimize a penalty score known as agony which penalized the edges from higher ranks to lower ranks.
We showed that the construction of such a label tree optimally is \np-hard, or even inapproximable when we limit the number of leaves.
Therefore, we presented a heuristic algorithm which runs in  $\bigO{(n + m)
\log n + \ell R}$, where $R = \sum_v \abs{L(v)}$ is the number of node-label
pairs in given graph, $\ell$ is the number of nodes in the resulting label
tree, and $n$ and $m$ are the number of nodes and edges respectively.  To
enforce the cardinality constraint for number of ranks $k$, we pruned the label
tree such that tree had only $k$ leaves, exploiting dynamic programming
techniques.

The synthetic experiments showed that our approach accurately recovers the
latent hierarchy. The experiments on  real-world networks confirmed that our
proposed label-driven algorithm achieved a good quality ranks
which can be explained by node labels. We discovered
hierarchies reasonably fast in practice. The notion of discovering hierarchies
in labeled networks opens up several lines of work. For example, 
instead of requiring that the nodes in each rank group match exactly to the
label tree we can require that only a large portion of the nodes match the label tree.


%% file: appendix.tex
\section{Proofs}

\begin{proof}[Proposition~\ref{prop:np}]
We will prove the hardness from $k$-\prbcover, a problem where we are given a
family of subsets $\fm{C}$ of a universe $U$ and we asked if there are at most $k$
sets covering $U$.

Assume that we are given an instance of \prbcover. Our graph is as follows: the
vertices $V$ consists of the items in $U = u_1, \ldots, u_n$ plus one additional
vertex $w$.  We connect each $u_i$ to $w$. The labels $L(u_i)$ are the indices
of the sets in which $u_i$ is included. Finally, $L(w) = \emptyset$.

We claim that there is a label tree $T$ with at most $k + 1$ leaves yielding
$\labelscore{G, T} = 0$ if and only if there is a $k$-cover.

Assume there is a cover $\fm{Z}$. Then let $T$ be a tree with $\abs{\fm{Z}}$ non-leaves.
Each of these nodes have a leaf as the left child, and one node has two leaves
as children. Each non-leaf has a label corresponding to a set in $\fm{Z}$
guiding the nodes in $G$ with the label to the left leaf.
The tree has at most $k + 1$ leaves, moreover the last leaf contains only
$w$ since it does not have any labels and $\fm{Z}$ is a cover. Consequently, all 
edges are forward, and $\labelscore{G, T} = 0$.

On the other hand, if $\labelscore{G, T} = 0$, then the last leaf contains only $w$
since otherwise there is a backward edge. Let $\fm{Z}$ be the sets corresponding
to the labels occurring in $T$. Then $\fm{Z}$ is a cover since otherwise there would
be a node with the same rank as $w$, violating the assumption. Moreover, $\abs{\fm{Z}} \leq k$
proving the claim.
\end{proof}

\begin{proof}[Proposition~\ref{prop:np2}]
We will prove the hardness from $k$-\prbcover.
Assume that we are given an instance of \prbcover:
a universe $U$ of $n$ items, a
family of $m$ subsets $\fm{C}$, and an integer $k$.

Our graph is as follows: the
vertices $V$ consists of the items in $U = u_1, \ldots, u_m$, $m$ vertices representing
the sets $S = s_1, \ldots, s_m$, 
and one additional
vertex $x$.  We connect each $u_i$ to $x$ with a weight $w(e) = m + k(k + 1)$.
We connect $x$ to each $s_i$ with a weight $w(e) = 1$.

The labels $L(u_i)$ are the indices of the sets in which $u_i$ is included. The
single label for $s_i$ is the index of the $i$th set.  Finally, $L(w) =
\emptyset$.

We claim that there is a label tree $T$ yielding
$\labelscore{G, T} \leq m + k(k + 1)/2$ if and only if there is a $k$-cover.

Assume there is a cover $\fm{Z}$. Then let $T$ be a tree with $\abs{\fm{Z}}$ non-leaves.
Each of these nodes have a leaf as the left child, and one node has two leaves
as children. Each non-leaf has a label corresponding to a set in $\fm{Z}$
guiding the nodes in $G$ with the label to the left leaf.
Since $\fm{Z}$ is a cover all 
edges $(u_i, x)$ are forward. Moreover, $k$ vertices in $S$ have ranks $1, \ldots, k$
and the remaining $m - k$ vertices in $S$ have the same rank as $x$, that is $k + 1$. Consequently,
\[
	\labelscore{G, T} = m + k(k + 1)/2\quad.
\]

On the other hand, if $\labelscore{G, T} \leq m + k(k + 1)/2$, then the last leaf contains only $x$
and vertices from $S$, since $w(u_i, x) > m + k(k + 1)/2$.

Let $\fm{Z}$ be the $o$ sets corresponding to the labels occurring in the path to
$x$ from the root, in order. Then $\fm{Z}$ is a cover since otherwise there is a
node in $U$ with the same rank as $x$, violating the assumption. 
Let $z_1, \ldots, z_o$ be the corresponding vertices in $S$.
Then $r(x) - r(z_i) \geq o - i + 1$ since there are at least $o - i$ leaves between $z_i$ and $x$.
Since $x$ has the largest rank, we also have $r(x) \geq r(s_i)$. Consequently,
\[
\begin{split}
	 m + \frac{k(k + 1)}{2} & \geq \labelscore{G, T} = \sum_{i = 1}^m 1 + r(x) - r(s_i) \\
	 & \geq m - o + \sum_{i = 1}^o (o - i + 2) = m + \frac{o(o + 1)}{2}, \\
\end{split}
\]
which shows that $o \leq k$, proving the result.
\end{proof}

\begin{proof}[Proposition~\ref{prop:time}]
$\algpartition(\alpha)$ first finds the best candidate by calling
repeatedly $\algsplit(\alpha, t)$,
and then performs the split by calling $\algleft(\alpha)$.

Since $\algsplit(\alpha, t)$ runs in $\bigO{\abs{V(\alpha, t)}}$ time,
the total time needed to find the candidate for splitting $\alpha$
is $\bigO{\sum_t \abs{V(\alpha, t)}} \subseteq \bigO{R}$.

Next we bound the time needed to update the counters during the split.
Let us define $i_{e\alpha} = 1$ if an edge $e$ is visited during $\algleft(\alpha)$,
and $i_{e\alpha} = 0$ otherwise.
Define also $i_{v\alpha} = 1$ if a node $v$ is visited during for-loop,
and $i_{v\alpha} = 0$ otherwise.

Write $n_\alpha = \sum i_{v\alpha}$ and $m_\alpha = \sum i_{e\alpha}$ to be the total number of nodes and edges visited during the split.

Let us define $c_\alpha = \abs{E(\alpha)} + \abs{V(\alpha)}$.
Testing whether $c_\beta \leq c_\gamma$ can be done in $\bigO{n_\alpha + m_\alpha}$, and
splitting $\alpha$ can be done in $\bigO{\abs{V(\alpha, t)} + n_\alpha + m_\alpha}$ time.

First note that $\abs{V(\alpha, t)} \leq R$.
The proposition then follows if we can prove that $\sum_\alpha n_\alpha + m_\alpha \in \bigO{(n + m) \log n}$.
Since 
\[
	\sum_\alpha n_\alpha + m_\alpha = \sum_v \sum_{\alpha} i_{v\alpha} + \sum_e \sum_{\alpha} i_{e\alpha},
\]
we will prove the claim by showing that $\sum_{\alpha} i_{e\alpha} \in \bigO{\log n}$
and  $\sum_{\alpha} i_{v\alpha} \in \bigO{\log n}$

Fix edge $e$, and let $\alpha$ and $\alpha'$ be two nodes for which $i_{e\alpha}
= i_{e\alpha'} = 1$. Then $\alpha'$ is a descendant of $\alpha$, or otherwise.
Assume the former. Then $\alpha'$ is either a child of $\alpha$ or
is a descendant of that child. Let us denote this child by $\beta$ and let $\gamma$
be the other child.

Since $e \in E(\alpha') \subseteq E(\beta)$, we have
$c_\beta \leq c_\gamma$ as otherwise $e$ is not visited
when $\alpha$ is split.
Consequently,
\[
	2 c_{\alpha'} \leq 2 c_{\beta} \leq c_\beta + c_\gamma \leq c_\alpha \quad.
\]

To conclude, let $\set{\alpha_j}$ be the nodes for which $i_{e\alpha_j} = 1$. We have shown that we can
safely assume that $\alpha_{j + 1}$ is a descendant of $\alpha_{j + 1}$. Moreover,
$c_{\alpha_j} \geq 2c_{\alpha_{j + 1}}$. Since $c_{\alpha_1} \leq m + n$,
there can be at most $\bigO{\log m} \subseteq \bigO{\log n}$ nodes.
The argument for $\sum_{\alpha} i_{v\alpha} \in \bigO{\log n}$ is similar.
\end{proof}

\section{Additional experiments}


\textbf{Synthetic data:}
As an additional experiment with synthetic data, our goal was to showcase how well 
the algorithm finds the ground truth for the dataset where there are no labels
that match the ranks exactly. Instead we considered a ground truth tree
given in Figure~\ref{fig:tree-gt}. 

Note that here
thick lines indicate the availability of the label and
dashed lines represent the unavailability.

\begin{figure}[ht!]
\begin{center}
\setlength{\tabcolsep}{0pt}
\begin{tabular}{ll}

\begin{tikzpicture}[baseline = 0pt,yscale=0.75, xscale=0.9]
\node (n0) at (-5, 1.5) {$\shortstack{L-6}$};

\node (n1) at (-4, 0) {$\shortstack{L-2}$};

\node (n18) at (-6, 0) {$\shortstack{ \color{purple} R-1}$};

\node (n2)  at (-2, -1.7) {$\shortstack{L-1}
 $};
\node (n3) at (-6, -1.7) {$\shortstack{L-4}
 $};

\node (n4) at (-3, -3.4) {$\shortstack{L-5}$};

\node (n10) at (-3.6, -5.0) {$ \shortstack{ \color{purple} R-5}$};

\node (n11) at (-2.4, -5.0) {$ \shortstack{ \color{purple} R-6}$};
\node (n16) at (-1.9, -6.4) {$ \shortstack{ \color{purple} R-7}$};
\node (n17) at (-1.1, -6.4) {$ \shortstack{ \color{purple} R-8}$};

\node (n5) at (-1, -3.4) {$\shortstack{L-3}$};
\node (n12) at (-.5, -5.0) {$ \shortstack{ \color{purple} R-9}$};
\node (n13) at (-1.5, -5.0) {$ \shortstack{  L-8}$};

\node (n6) at (-5, -3.4) {$\shortstack{L-7}$};
\node (n7) at (-7, -3.4) {$\shortstack{ \color{purple} R-2}$};

\node (n8) at (-5.5, -5) {$ \shortstack{ \color{purple} R-3}$};
\node (n9) at (-4.5, -5) {$ \shortstack{ \color{purple} R-4}$};

\draw[yafcolor3, ->, >=latex, thick, dashed, in=90, out=-90] (n1) edge (n2);
\draw[yafcolor3, ->, >=latex, thick, in=90, out=-90] (n1) edge (n3);

\draw[yafcolor3, ->, >=latex, thick, dashed, in=90, out=-90] (n2) edge (n4);
\draw[yafcolor3, ->, >=latex, thick, in=90, out=-90] (n2) edge (n5);

\draw[yafcolor3, ->, >=latex, thick, dashed, in=90, out=-90] (n3) edge (n6);
\draw[yafcolor3, ->, >=latex, thick, in=90, out=-90] (n3) edge (n7);

\draw[yafcolor3, ->, >=latex, thick, in=90, out=-90] (n4) edge (n10);
\draw[yafcolor3, ->, >=latex, thick, dashed, in=90, out=-90] (n4) edge (n11);

\draw[yafcolor3, ->, >=latex, thick, in=90, out=-90] (n5) edge (n12);
\draw[yafcolor3, ->, >=latex, thick, dashed, in=90, out=-90] (n5) edge (n13);

\draw[yafcolor3, ->, >=latex, thick, in=90, out=-90] (n6) edge (n8);
\draw[yafcolor3, ->, >=latex, thick, dashed, in=90, out=-90] (n6) edge (n9);

\draw[yafcolor3, ->, >=latex, thick, dashed, in=90, out=-90] (n13) edge (n16);
\draw[yafcolor3, ->, >=latex, thick, in=90, out=-90] (n13) edge (n17);

\draw[yafcolor3, ->, >=latex, thick, in=90, out=-90] (n0) edge (n18);
\draw[yafcolor3, ->, >=latex, thick, dashed, in=90, out=-90] (n0) edge (n1);

\end{tikzpicture}
\end{tabular}
\caption{Discovered label tree obtained from noiseless network~($\theta = 0$ and  $\mu = 0$).
}
\label{fig:tree-noiseless}
\end{center}
\end{figure}
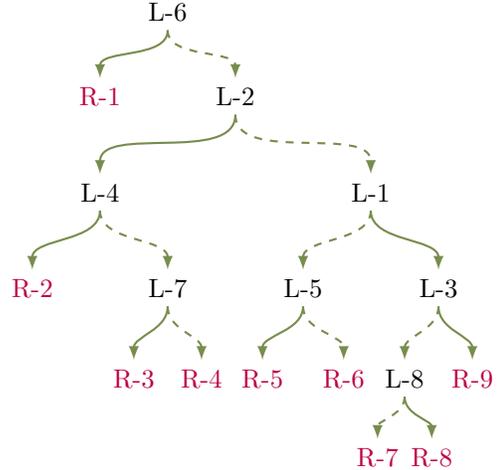

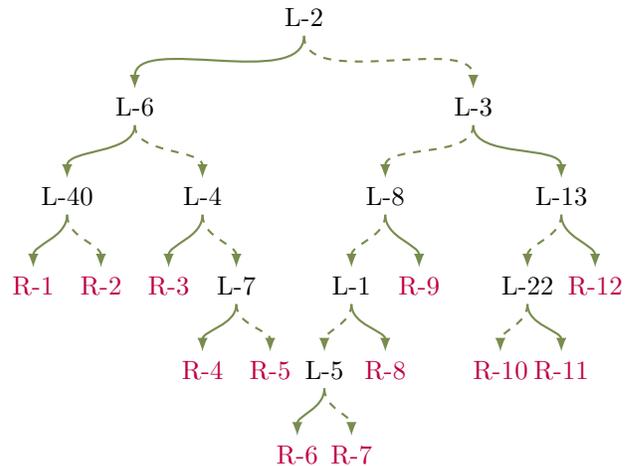
\begin{figure}[ht!]
\begin{center}
\setlength{\tabcolsep}{0pt}
\begin{tabular}{ll}

\begin{tikzpicture}[baseline = 0pt,yscale=0.7, xscale=0.9]
\node (n0) at (-6.5, 1.7) {$\shortstack{L-2}$};

\node (n1) at (-4, 0) {$\shortstack{L-3}$};

\node (n18) at (-9, 0) {$\shortstack{  L-6}$};
\node (n19) at (-10, -1.7) {$\shortstack{  L-40}$};
\node (n20) at (-8, -1.7) {$\shortstack{  L-4}$};

\node (n21) at (-10.5, -3.4) {$\shortstack{ \color{purple} R-1}$};
\node (n22) at (-9.5, -3.4) {$\shortstack{ \color{purple} R-2}$};

\node (n23) at (-8.5, -3.4) {$\shortstack{ \color{purple} R-3}$};
\node (n24) at (-7.5, -3.4) {$\shortstack{  L-7}$};

\node (n25) at (-8, -5) {$ \shortstack{ \color{purple} R-4}$};
\node (n26) at (-7, -5) {$ \shortstack{ \color{purple} R-5}$};

\node (n2)  at (-2.7, -1.7) {$\shortstack{L-13}
 $};
\node (n3) at (-5.3, -1.7) {$\shortstack{L-8}
 $};

\node (n4) at (-3.2, -3.4) {$\shortstack{L-22}$};

\node (n10) at (-3.6, -5.0) {$ \shortstack{ \color{purple} R-10}$};
\node (n11) at (-2.7, -5.0) {$ \shortstack{ \color{purple} R-11}$};

\node (n5) at (-2.2, -3.4) {$\shortstack{\color{purple} R-12}$};

\node (n6) at (-4.8, -3.4) {$\shortstack{ \color{purple} R-9}$};
\node (n7) at (-5.8, -3.4) {$\shortstack{ L-1}$};

\node (n8) at (-6.2, -5) {$ \shortstack{  L-5}$};
\node (n27) at (-6.6, -6.6) {$ \shortstack{ \color{purple} R-6}$};
\node (n28) at (-5.8, -6.6) {$ \shortstack{ \color{purple} R-7}$};
\node (n9) at (-5.3, -5) {$ \shortstack{ \color{purple} R-8}$};

\draw[yafcolor3, ->, >=latex, thick, in=90, out=-90] (n1) edge (n2);
\draw[yafcolor3, ->, >=latex, thick, dashed, in=90, out=-90] (n1) edge (n3);

\draw[yafcolor3, ->, >=latex, thick, dashed, in=90, out=-90] (n2) edge (n4);
\draw[yafcolor3, ->, >=latex, thick, in=90, out=-90] (n2) edge (n5);

\draw[yafcolor3, ->, >=latex, thick, in=90, out=-90] (n3) edge (n6);
\draw[yafcolor3, ->, >=latex, thick, dashed, in=90, out=-90] (n3) edge (n7);

\draw[yafcolor3, ->, >=latex, thick, dashed, in=90, out=-90] (n4) edge (n10);
\draw[yafcolor3, ->, >=latex, thick, in=90, out=-90] (n4) edge (n11);

\draw[yafcolor3, ->, >=latex, thick, dashed, in=90, out=-90] (n7) edge (n8);
\draw[yafcolor3, ->, >=latex, thick, in=90, out=-90] (n7) edge (n9);

\draw[yafcolor3, ->, >=latex, thick, in=90, out=-90] (n0) edge (n18);
\draw[yafcolor3, ->, >=latex, thick, dashed, in=90, out=-90] (n0) edge (n1);

\draw[yafcolor3, ->, >=latex, thick, in=90, out=-90] (n18) edge (n19);
\draw[yafcolor3, ->, >=latex, thick, dashed, in=90, out=-90] (n18) edge (n20);

\draw[yafcolor3, ->, >=latex, thick, in=90, out=-90] (n19) edge (n21);
\draw[yafcolor3, ->, >=latex, thick, dashed, in=90, out=-90] (n19) edge (n22);

\draw[yafcolor3, ->, >=latex, thick, in=90, out=-90] (n20) edge (n23);
\draw[yafcolor3, ->, >=latex, thick, dashed, in=90, out=-90] (n20) edge (n24);

\draw[yafcolor3, ->, >=latex, thick, in=90, out=-90] (n24) edge (n25);
\draw[yafcolor3, ->, >=latex, thick, dashed, in=90, out=-90] (n24) edge (n26);

\draw[yafcolor3, ->, >=latex, thick, in=90, out=-90] (n8) edge (n27);
\draw[yafcolor3, ->, >=latex, thick, dashed, in=90, out=-90] (n8) edge (n28);

\end{tikzpicture}
\end{tabular}
\caption{Discovered label tree obtained from noisy network~($\theta = 0.3$ and  $\mu = 0.3$).
}
\label{fig:tree-noisy}
\end{center}
\end{figure}

We generated a new dataset with $n =
9\,000$, $m = 9\,600$, $h = 9$, and $\eta = 0.75$, in accordance with the true
label-sets derived from the given tree.  The discovered label trees with
noiseless~($\theta = 0$ and  $\mu = 0$)
and
noisy~($\theta = 0.3$ and  $\mu = 0.3$)
parameters are shown in Figure~\ref{fig:tree-noiseless} and
Figure~\ref{fig:tree-noisy} respectively.
When compared to the ground truth,
we achieved the value $1$ for
Kendall's $\tau$ measure without the noise and the
$0.928$ with the noise. Note that the
discovered tree in Figure~\ref{fig:tree-noiseless} is not the same as the ground truth tree
but it will yield the same ranking.


\begin{figure}[ht!]
\begin{center}
\setlength{\tabcolsep}{0pt}
\begin{tabular}{ll}

\begin{tikzpicture}[baseline = 0pt, yscale=0.7, xscale=1]
\node (n1) at (0, 0) {$\shortstack{L-1}$};

\node (n2)  at (-2, -1.7) {$\shortstack{L-2}
 $};
\node (n3) at (2, -1.7) {$\shortstack{L-3}
 $};

\node (n4) at (-3, -3.4) {$\shortstack{L-4}$};

\node (n10) at (-3.7, -5.0) {$ \shortstack{ L-6}$};
\node (n14) at (-4.0, -6.4) {$ \shortstack{ \color{purple} R-1}$};
\node (n15) at (-3.4, -6.4) {$ \shortstack{ \color{purple} R-2}$};

\node (n11) at (-2.3, -5.0) {$ \shortstack{ L-7}$};
\node (n16) at (-2.6, -6.4) {$ \shortstack{ \color{purple} R-3}$};
\node (n17) at (-2.0, -6.4) {$ \shortstack{ \color{purple} R-4}$};

\node (n5) at (-1, -3.4) {$\shortstack{L-5}$};
\node (n12) at (-.5, -5.0) {$ \shortstack{ \color{purple} R-6}$};
\node (n13) at (-1.5, -5.0) {$ \shortstack{ \color{purple} R-5}$};

\node (n6) at (1, -3.4) {$\shortstack{L-8}$};
\node (n7) at (3, -3.4) {$\shortstack{ \color{purple} R-9}$};

\node (n8) at (0.5, -5) {$ \shortstack{ \color{purple} R-7}$};
\node (n9) at (1.5, -5) {$ \shortstack{ \color{purple} R-8}$};

\draw[yafcolor3, ->, >=latex, thick, dashed, in=90, out=-90] (n1) edge (n2);
\draw[yafcolor3, ->, >=latex, thick, in=90, out=-90] (n1) edge (n3);

\draw[yafcolor3, ->, >=latex, thick, in=90, out=-90] (n2) edge (n4);
\draw[yafcolor3, ->, >=latex, thick, dashed, in=90, out=-90] (n2) edge (n5);

\draw[yafcolor3, ->, >=latex, thick, dashed, in=90, out=-90] (n3) edge (n6);
\draw[yafcolor3, ->, >=latex, thick, in=90, out=-90] (n3) edge (n7);

\draw[yafcolor3, ->, >=latex, thick, in=90, out=-90] (n4) edge (n10);
\draw[yafcolor3, ->, >=latex, thick, dashed, in=90, out=-90] (n4) edge (n11);

\draw[yafcolor3, ->, >=latex, thick, dashed, in=90, out=-90] (n5) edge (n12);
\draw[yafcolor3, ->, >=latex, thick, in=90, out=-90] (n5) edge (n13);

\draw[yafcolor3, ->, >=latex, thick, dashed, in=90, out=-90] (n6) edge (n8);
\draw[yafcolor3, ->, >=latex, thick, in=90, out=-90] (n6) edge (n9);

\draw[yafcolor3, ->, >=latex, thick, in=90, out=-90] (n10) edge (n14);
\draw[yafcolor3, ->, >=latex, thick, dashed, in=90, out=-90] (n10) edge (n15);

\draw[yafcolor3, ->, >=latex, thick, in=90, out=-90] (n11) edge (n16);
\draw[yafcolor3, ->, >=latex, thick, dashed, in=90, out=-90] (n11) edge (n17);

\end{tikzpicture}
\end{tabular}
\caption{Ground truth label tree.}
\label{fig:tree-gt}
\end{center}
\end{figure}
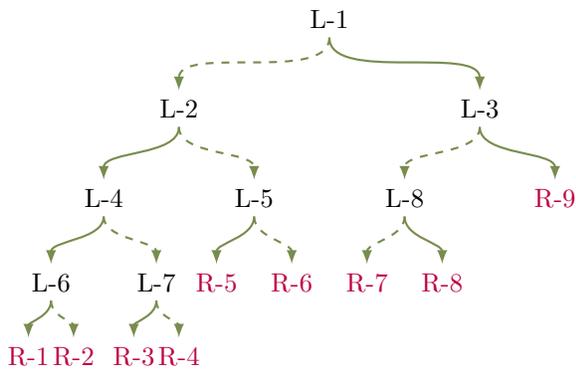




\textbf{Real-world data:}
Let us look at the trees obtained \dtname{Patent-citation} dataset together with their labels. This tree shows $7$ levels. The label for the very first  partition has become a patent class \emph{** Classification Undetermined **}. Then, the tree has been further split to the left side, based on the label $Computers \& communications$ which is a sub category.

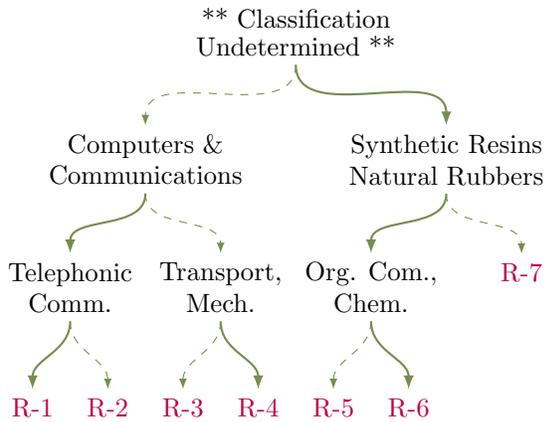
\begin{figure}[ht!]
\begin{center}
\setlength{\tabcolsep}{0pt}
\begin{tabular}{ll}

\begin{tikzpicture}[baseline = 0pt]
\node (n1) at (0, 0) {$\shortstack{** Classification  
\\Undetermined **}$};

\node (n2)  at (-2, -1.7) {$\shortstack{Computers  \& 
\\Communications}
 $};
\node (n3) at (2, -1.7) {$\shortstack{Synthetic Resins \\Natural Rubbers}
 $};

\node (n4) at (-3, -3.4) {$\shortstack{Telephonic \\Comm.}$};
\node (n10) at (-3.5, -5.0) {$ \shortstack{ \color{purple} R-1}$};
\node (n11) at (-2.5, -5.0) {$ \shortstack{ \color{purple} R-2}$};

\node (n5) at (-1, -3.4) {$\shortstack{Transport, \\Mech.}$};
\node (n12) at (-.5, -5.0) {$ \shortstack{ \color{purple} R-4}$};
\node (n13) at (-1.5, -5.0) {$ \shortstack{ \color{purple} R-3}$};

\node (n6) at (1, -3.4) {$\shortstack{Org. Com., \\Chem.}$};
\node (n7) at (3, -3.2) {$\shortstack{ \color{purple} R-7}$};
\node (n8) at (0.5, -5) {$ \shortstack{ \color{purple} R-5}$};
\node (n9) at (1.5, -5) {$ \shortstack{ \color{purple} R-6}$};

\draw[yafcolor3, ->, >=latex, dashed, in=90, out=-90] (n1) edge (n2);
\draw[yafcolor3, ->, >=latex, thick, in=90, out=-90] (n1) edge (n3);

\draw[yafcolor3, ->, >=latex, thick, in=90, out=-90] (n2) edge (n4);
\draw[yafcolor3, ->, >=latex, dashed, in=90, out=-90] (n2) edge (n5);

\draw[yafcolor3, ->, >=latex, thick, in=90, out=-90] (n3) edge (n6);
\draw[yafcolor3, ->, >=latex, dashed, in=90, out=-90] (n3) edge (n7);

\draw[yafcolor3, ->, >=latex, thick, in=90, out=-90] (n4) edge (n10);
\draw[yafcolor3, ->, >=latex, dashed, in=90, out=-90] (n4) edge (n11);

\draw[yafcolor3, ->, >=latex, thick, in=90, out=-90] (n5) edge (n12);
\draw[yafcolor3, ->, >=latex, dashed, in=90, out=-90] (n5) edge (n13);

\draw[yafcolor3, ->, >=latex, dashed, in=90, out=-90] (n6) edge (n8);
\draw[yafcolor3, ->, >=latex, thick, in=90, out=-90] (n6) edge (n9);

\end{tikzpicture}
\end{tabular}
\caption{Label tree for \dtname{Patent-citation} dataset using Algorithm \ref{alg:parttion-fast}. Solid lines indicate the branch for the nodes with the corresponding label.}
\label{fig:tree-patent}
\end{center}
\end{figure}

\begin{figure}[ht!]
\begin{center}
\setlength{\tabcolsep}{0pt}
\begin{tabular}{ll}

\begin{tikzpicture}[baseline = 0pt, yscale=0.7, xscale=1]
\node (n1) at (0, 0) {$\shortstack{G-85}$};

\node (n2)  at (-2, -1.7) {$\shortstack{G-134}
 $};
\node (n3) at (2, -1.7) {$\shortstack{G-104}
 $};
 \node (n6) at (1, -3.4) {$\shortstack{\color{purple} R-4}$};
 \node (n7) at (3, -3.4) {$\shortstack{\color{purple} R-5}$};

\node (n4) at (-3, -3.4) {$\shortstack{\color{purple} R-1}$};

\node (n5) at (-1, -3.4) {$\shortstack{G-150}$};
\node (n12) at (0, -5.0) {$ \shortstack{ \color{purple} R-3}$};
\node (n13) at (-2, -5.0) {$ \shortstack{ \color{purple} R-2}$};

\draw[yafcolor3, ->, >=latex, thick, in=90, out=-90] (n1) edge (n2);
\draw[yafcolor3, ->, >=latex, dashed, in=90, out=-90] (n1) edge (n3);

\draw[yafcolor3, ->, >=latex, thick, in=90, out=-90] (n2) edge (n4);
\draw[yafcolor3, ->, >=latex, dashed, in=90, out=-90] (n2) edge (n5);

\draw[yafcolor3, ->, >=latex, dashed, in=90, out=-90] (n3) edge (n6);
\draw[yafcolor3, ->, >=latex, thick, in=90, out=-90] (n3) edge (n7);

\draw[yafcolor3, ->, >=latex, dashed, in=90, out=-90] (n5) edge (n12);
\draw[yafcolor3, ->, >=latex, thick, in=90, out=-90] (n5) edge (n13);

\end{tikzpicture}
\end{tabular}
\caption{Label tree for \dtname{Flickr} dataset using Algorithm \ref{alg:parttion-fast}. Solid lines indicate the branch for the nodes with the corresponding label.}
\label{fig:tree-flickr}
\end{center}
\end{figure}
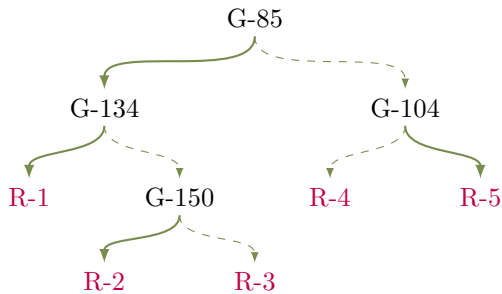

Figure~\ref{fig:tree-flickr} shows the label tree obtained with \dtname{Flickr} dataset. This tree shows $5$ hierarchy levels in total. We can observe that the nodes with label \emph{G-104} is ranked higher with respect to the nodes which include \emph{G-134} as its label.

\begin{figure}[ht!]
\begin{center}
\setlength{\tabcolsep}{0pt}
\begin{tabular}{ll}

\begin{tikzpicture}[baseline = 0pt, yscale=0.7, xscale=1]
\node (n1) at (0, 0) {$\shortstack{Case based}$};

\node (n2)  at (-2, -1.7) {$\shortstack{Theory}
 $};
\node (n3) at (2, -1.7) {$\shortstack{\color{purple} Rank 3}
 $};

\node (n4) at (-3, -3.4) {$\shortstack{\color{purple} Rank 1}$};
\node (n5) at (-1, -3.4) {$\shortstack{\color{purple} Rank 2}$};

\draw[yafcolor3, ->, >=latex, dashed, in=90, out=-90] (n1) edge (n2);
\draw[yafcolor3, ->, >=latex, thick, in=90, out=-90] (n1) edge (n3);

\draw[yafcolor3, ->, >=latex, thick, in=90, out=-90] (n2) edge (n4);
\draw[yafcolor3, ->, >=latex, dashed, in=90, out=-90] (n2) edge (n5);

\end{tikzpicture}
\end{tabular}
\caption{Label tree for \dtname{Cora} dataset using Algorithm \ref{alg:parttion-fast}. Solid lines indicate the branch for the nodes with the corresponding label.}
\label{fig:tree-cora}
\end{center}
\end{figure}
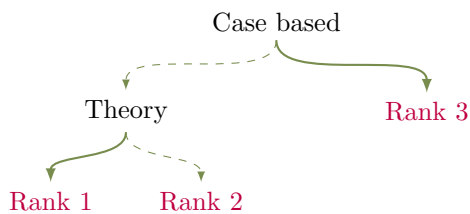

Let us look at Figure~\ref{fig:tree-cora} which demonstrates the label tree obtained for \dtname{Cora} dataset. This tree shows $3$ hierarchy levels in total. Based on this tree, we can argue that the  case-based papers are ranked higher with respect to the theoretical papers.

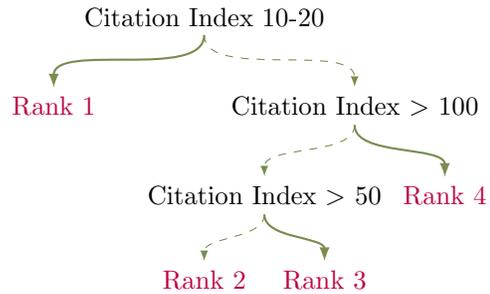
\begin{figure}[ht!]
\begin{center}
\setlength{\tabcolsep}{0pt}
\begin{tabular}{ll}

\begin{tikzpicture}[baseline = 0pt, yscale=0.7, xscale=1]
\node (n1) at (0, 0) {$\shortstack{Citation Index 10-20}$};

\node (n2)  at (-2, -1.7) {$\shortstack{\color{purple} Rank 1}
 $};
\node (n3) at (2, -1.7) {$\shortstack{Citation Index $>$ 100}$};

\node (n4) at (0.8, -3.4) {$\shortstack{Citation Index $>$ 50}$};
\node (n6) at (0.0, -5.0) {$\shortstack{\color{purple} Rank 2}$};
\node (n7) at (1.6, -5.0) {$\shortstack{\color{purple} Rank 3}$};

\node (n5) at (3.2, -3.4) {$\shortstack{\color{purple} Rank 4}$};

\draw[yafcolor3, ->, >=latex, thick, in=90, out=-90] (n1) edge (n2);
\draw[yafcolor3, ->, >=latex, dashed, in=90, out=-90] (n1) edge (n3);

\draw[yafcolor3, ->, >=latex, dashed, in=90, out=-90] (n3) edge (n4);
\draw[yafcolor3, ->, >=latex, thick, in=90, out=-90] (n3) edge (n5);

\draw[yafcolor3, ->, >=latex, dashed, in=90, out=-90] (n4) edge (n6);
\draw[yafcolor3, ->, >=latex, thick, in=90, out=-90] (n4) edge (n7);

\end{tikzpicture}
\end{tabular}
\caption{Label tree for \dtname{EIES} dataset using Algorithm \ref{alg:parttion-fast}. Solid lines indicate the branch for the nodes with the corresponding label.}
\label{fig:tree-opal}
\end{center}
\end{figure}

Next we observe Figure~\ref{fig:tree-opal} which provides the label tree constructed using \dtname{EIES} dataset. This tree shows $4$ hierarchy levels in total. Based on the interactions between researchers, our algorithm classifies the researchers with citation index greater than $100$ in higher rank with respect to the researchers with citation index between $10$ and $20$.

\begin{figure}[ht!]
\begin{center}
\setlength{\tabcolsep}{0pt}
\begin{tabular}{ll}

\begin{tikzpicture}[baseline = 0pt, yscale=0.7, xscale=1]
\node (n1) at (0, 0) {$\shortstack{Agents}$};

\node (n2)  at (-2, -1.7) {$\shortstack{Machine Learning}
 $};
\node (n3) at (2, -1.7) {$\shortstack{\color{purple} Rank 3}
 $};

\node (n4) at (-3, -3.4) {$\shortstack{\color{purple} Rank 1}$};
\node (n5) at (-1, -3.4) {$\shortstack{\color{purple} Rank 2}$};

\draw[yafcolor3, ->, >=latex, dashed, in=90, out=-90] (n1) edge (n2);
\draw[yafcolor3, ->, >=latex, thick, in=90, out=-90] (n1) edge (n3);

\draw[yafcolor3, ->, >=latex, thick, in=90, out=-90] (n2) edge (n4);
\draw[yafcolor3, ->, >=latex, dashed, in=90, out=-90] (n2) edge (n5);

\end{tikzpicture}
\end{tabular}
\caption{Label tree for \dtname{Citeseer} dataset using Algorithm \ref{alg:parttion-fast}. Solid lines indicate the branch for the nodes with the corresponding label.}
\label{fig:tree-citeseer}
\end{center}
\end{figure}
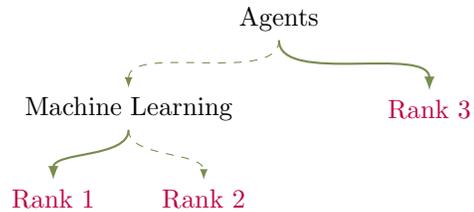

Next, Figure~\ref{fig:tree-citeseer} demonstrates the label tree obtained using \dtname{Citeseer} dataset. This tree shows $3$ hierarchy levels in total. We can see that
the papers in \emph{Agents} category are cited more than the papers which are categorized under \emph{Machine Learning}.

\begin{figure}[ht!]
\begin{center}
\setlength{\tabcolsep}{0pt}
\begin{tabular}{ll}

\begin{tikzpicture}[baseline = 0pt, yscale=0.7, xscale=1]
\node (n1) at (0, 0) {$\shortstack{Biology-1}$};

\node (n2)  at (-2, -1.7) {$\shortstack{\color{purple} Rank 1}
 $};
\node (n3) at (2, -1.7) {$\shortstack{Mathematics and Physics}$};

\node (n4) at (0.8, -3.4) {$\shortstack{\color{purple} Rank 2}$};

\node (n5) at (3.2, -3.4) {$\shortstack{\color{purple} Rank 3}$};

\draw[yafcolor3, ->, >=latex, thick, in=90, out=-90] (n1) edge (n2);
\draw[yafcolor3, ->, >=latex, dashed, in=90, out=-90] (n1) edge (n3);

\draw[yafcolor3, ->, >=latex, dashed, in=90, out=-90] (n3) edge (n4);
\draw[yafcolor3, ->, >=latex, thick, in=90, out=-90] (n3) edge (n5);

\end{tikzpicture}
\end{tabular}
\caption{Label tree for \dtname{Student} dataset using Algorithm \ref{alg:parttion-fast}. Solid lines indicate the branch for the nodes with the corresponding label.}
\label{fig:tree-student}
\end{center}
\end{figure}
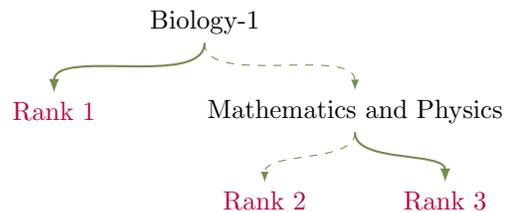

Next we focus on Figure~\ref{fig:tree-student} which shows the label tree obtained \dtname{Student} dataset. This tree shows $3$ hierarchy levels in total. Based on the contacts between students, it is evident that  mathematics and physics class students are highly sorted than biology class students.

%% file: paper.bbl
\begin{thebibliography}{19}
\providecommand{\natexlab}[1]{#1}
\providecommand{\url}[1]{\texttt{#1}}
\expandafter\ifx\csname urlstyle\endcsname\relax
  \providecommand{\doi}[1]{doi: #1}\else
  \providecommand{\doi}{doi: \begingroup \urlstyle{rm}\Url}\fi

\bibitem[Bai et~al.(2020)Bai, Ravi, and Davidson]{bai2020towards}
Z.~Bai, S.~Ravi, and I.~Davidson.
\newblock Towards description of block model on graph.
\newblock In \emph{ECMLPKDD}, pages 37--53, 2020.

\bibitem[Bothorel et~al.(2015)Bothorel, Cruz, Magnani, and Micenkova]{bothorel2015clustering}
C.~Bothorel, J.~D. Cruz, M.~Magnani, and B.~Micenkova.
\newblock Clustering attributed graphs: models, measures and methods.
\newblock \emph{Network Science}, 3\penalty0 (3):\penalty0 408--444, 2015.

\bibitem[Dinur and Safra(2005)]{dinur2005hardness}
I.~Dinur and S.~Safra.
\newblock On the hardness of approximating minimum vertex cover.
\newblock \emph{Annals of mathematics}, pages 439--485, 2005.

\bibitem[Elo(1978)]{elo1978rating}
A.~E. Elo.
\newblock \emph{The rating of chessplayers, past and present}.
\newblock Arco Pub., New York, 1978.

\bibitem[Even et~al.(1998)Even, Schieber, Sudan, et~al.]{even1998approximating}
G.~Even, B.~Schieber, M.~Sudan, et~al.
\newblock Approximating minimum feedback sets and multicuts in directed graphs.
\newblock \emph{Algorithmica}, 20\penalty0 (2):\penalty0 151--174, 1998.

\bibitem[Falih et~al.(2018)Falih, Grozavu, Kanawati, and Bennani]{falih2018community}
I.~Falih, N.~Grozavu, R.~Kanawati, and Y.~Bennani.
\newblock Community detection in attributed network.
\newblock In \emph{WWW}, pages 1299--1306, 2018.

\bibitem[Galbrun et~al.(2014)Galbrun, Gionis, and Tatti]{galbrun2014overlapping}
E.~Galbrun, A.~Gionis, and N.~Tatti.
\newblock Overlapping community detection in labeled graphs.
\newblock \emph{DMKD}, 28\penalty0 (5):\penalty0 1586--1610, 2014.

\bibitem[Gupte et~al.(2011)Gupte, Shankar, Li, Muthukrishnan, and Iftode]{gupte2011finding}
M.~Gupte, P.~Shankar, J.~Li, S.~Muthukrishnan, and L.~Iftode.
\newblock Finding hierarchy in directed online social networks.
\newblock In \emph{WWW}, pages 557--566, 2011.

\bibitem[Jameson et~al.(1999)Jameson, Appleby, and Freeman]{jameson1999finding}
K.~A. Jameson, M.~C. Appleby, and L.~C. Freeman.
\newblock Finding an appropriate order for a hierarchy based on probabilistic dominance.
\newblock \emph{Animal behaviour}, 57\penalty0 (5):\penalty0 991--998, 1999.

\bibitem[Kleinberg(1999)]{kleinberg1999authoritative}
J.~M. Kleinberg.
\newblock Authoritative sources in a hyperlinked environment.
\newblock \emph{JACM}, 46\penalty0 (5):\penalty0 604--632, 1999.

\bibitem[Lu et~al.(2018)Lu, Chen, and Zhang]{UDMS_dataset}
J.~Lu, J.~Chen, and C.~Zhang.
\newblock Helsinki {M}ulti-{M}odel {D}ata {R}epository.
\newblock \url{https://www.helsinki.fi/en/researchgroups/unified-database-management-systems-udbms/datasets/patent-dataset}, 2018.

\bibitem[Maiya and Berger-Wolf(2009)]{maiya2009inferring}
A.~S. Maiya and T.~Y. Berger-Wolf.
\newblock Inferring the maximum likelihood hierarchy in social networks.
\newblock In \emph{CSE}, volume~4, pages 245--250, 2009.

\bibitem[Memon et~al.(2008)Memon, Larsen, Hicks, and Harkiolakis]{memon2008retracted}
N.~Memon, H.~L. Larsen, D.~L. Hicks, and N.~Harkiolakis.
\newblock Retracted: detecting hidden hierarchy in terrorist networks: some case studies.
\newblock In \emph{ISI}, pages 477--489, 2008.

\bibitem[Neumann et~al.(2018)Neumann, Ritter, and Budhathoki]{Neumann2018RankingTT}
S.~Neumann, J.~Ritter, and K.~Budhathoki.
\newblock Ranking the teams in european football leagues with agony.
\newblock In \emph{MLSA@ECMLPKDD}, 2018.

\bibitem[Pool et~al.(2014)Pool, Bonchi, and Leeuwen]{pool2014description}
S.~Pool, F.~Bonchi, and M.~v. Leeuwen.
\newblock Description-driven community detection.
\newblock \emph{TIST}, 5\penalty0 (2):\penalty0 1--28, 2014.

\bibitem[Rossi and Ahmed(2015)]{nr2015}
R.~A. Rossi and N.~K. Ahmed.
\newblock The network data repository with interactive graph analytics and visualization.
\newblock In \emph{AAAI}, 2015.
\newblock URL \url{https://networkrepository.com}.

\bibitem[Rowe et~al.(2007)Rowe, Creamer, Hershkop, and Stolfo]{rowe2007automated}
R.~Rowe, G.~Creamer, S.~Hershkop, and S.~J. Stolfo.
\newblock Automated social hierarchy detection through email network analysis.
\newblock In \emph{WebKDD/SNA-KDD}, pages 109--117, 2007.

\bibitem[Tang et~al.(2008)Tang, Zhang, Yao, Li, Zhang, and Su]{tang2008arnetminer}
J.~Tang, J.~Zhang, L.~Yao, J.~Li, L.~Zhang, and Z.~Su.
\newblock Arnetminer: extraction and mining of academic social networks.
\newblock In \emph{KDD}, pages 990--998, 2008.

\bibitem[Tatti(2017)]{nikolaj2017tiers}
N.~Tatti.
\newblock Tiers for peers: a practical algorithm for discovering hierarchy in weighted networks.
\newblock \emph{DMKD}, 31\penalty0 (3):\penalty0 702--738, 2017.

\end{thebibliography}
